# $\eta$–$\eta'$ mixing and its application in the $B^+$, $D^+$, $D_s^+ \to \eta^{(\prime)}\ell^+\nu_\ell$ decays

Yu-Jie Zhang,[*] Xing-Gang Wu,[†] Dan-Dan Hu,[‡] and Long Zeng[§]
*Department of Physics, Chongqing Key Laboratory for Strongly Coupled Physics, Chongqing University, Chongqing 401331, People's Republic of China*

In this paper, we take into account the intrinsic charm and gluonic contents into the $\eta - \eta'$ mixing scheme and formulate the tetramixing $\eta - \eta' - G - \eta_c$ to study the mixing properties of $\eta^{(\prime)}$ mesons. Using the newly derived mixing parameters, we calculate the transition form factors (TFFs) of $B^+/D^+/D_s^+ \to \eta^{(\prime)}$ within the quantum chromodynamics (QCD) light-cone sum rules up to next-to-leading order QCD corrections and twist-4 contributions. Using the extrapolated TFFs, we then calculate the decay widths and branching fractions of the semileptonic decays $B^+/D^+/D_s^+ \to \eta^{(\prime)}\ell^+\nu_\ell$. Our results are consistent with the recent Belle and BES-III measurements within reasonable errors.

## I. INTRODUCTION

The large branching fraction for the inclusive production of high-momentum $\eta'$ meson in $B$ meson decays presented by the CLEO Collaboration [1] has attracted considerable attention to explain the dynamical origin of this enhancement. Theoretically, there are two widely discussed mechanisms for this large signal, one is attributed to that the $\eta'$ meson has a large anomalously coupling to the gluonic field configurations [2–4] and the other suggests that the $\eta'$ meson has a large $c\bar{c}$ content [5,6]. Experimentally, the gluonic contents of the $\eta^{(\prime)}$ mesons have been inferred from the KLOE data on the radiative decays $P \to \gamma V$ and $V \to \gamma P$ [7,8] and from the BES data on the charmonium decays $J/\Psi \to VP$ [9,10], where $P$ stands for a pseudoscalar, $A$ for an axial-vector, and $V$ for a vector current.

The gluonic and the intrinsic charm compositions in the light pseudoscalar mesons, namely $\eta$ and $\eta'$, could have a pivotal influence on numerous hadronic processes and have emerged as one of the topics of significant interest. The connection of $\eta^{(\prime)}$ mesons to the gluon is important to investigate the possible glueball candidates [11]. For example, Ball *et al.* [12] investigated the gluonic content in the $\eta^{(\prime)}$ mesons phenomenologically, and they suggested that the nonleptonic $D_s$ decay channels such as $D_s \to \eta'\rho$ and $D_s \to \eta\rho$ can be used to search for possible glueball states. Experimentally, a large amount of experimental data has been accumulated by the *BABAR* and Belle Collaborations regarding the $B$, $D$, and $D_s$ mesons' semileptonic decays into pseudoscalar $\eta^{(\prime)}$ mesons, which provides a good platform for studying the gluonic and charm contents in $\eta^{(\prime)}$ mesons. All those decays are important for testing and understanding the Standard Model (SM) flavor interactions, particular for understanding the quantum chromodynamics (QCD) dynamics in the flavor physics as well as the flavor mixing.

The glueball can mix with mesons of the same quantum numbers made of quarks. For the isosinglet pseudoscalars, $\eta(547)$, $\eta'(958)$, $\eta(1295)$, $\eta(1405)$, $\eta(1475)$, $\eta(1760)$, and $\eta(2225)$ are listed in Particle Data Group (PDG) [13]. In addition, $X(2370)$ has the same quantum numbers as isosinglet pseudoscalar mesons and may also participate in the relevant mixing processes. Some of these mesons even have a more complicated internal substructure and many contain quark-antiquark, tetraquarks, and molecules, etc. Considering that for the weak decays of $B$, $D$, and $D_s$ mesons studied here, the phase space strongly favors the production of the lightest physical states. So we adopt the mixture of the $\eta(547)$ and $\eta'(958)$ with $\eta_c$ and glueball to do our discussion. In principle the decay processes of $B$, $D$, and $D_s$ mesons does involve all of these etas that can have structures beyond quark-antiquark [14–20].

It has long been realized that the $\eta - \eta'$ mixing is caused by the axial $U(1)_A$ anomaly [21], i.e., the SU(3) breaking effects that are sizable and have a nontrivial structure [22], which have not been reliably calculated yet. In exact SU(3) flavor limit, the $\eta$ meson can be treated as a pure flavor-octet state labeled by $|\eta_8\rangle$ and the $\eta'$ meson is a pure flavor-singlet state labeled by $|\eta_1\rangle$ [23]. But due to the fact that the mass of strange quark is much larger than that of the $u/d$-quark, the SU(3) flavor symmetry will be explicitly destroyed, resulting as the mixing of the flavor-singlet state with the flavor-octet state. Moreover, the flavor-singlet state can also be mixed with a heavier intrinsic quark-antiquark state and a gluonic state, which will produce a large gluonic admixture in $\eta'$ meson and almost negligible ones in $\eta$ meson. Currently, the gluonic components within $\eta$ and $\eta'$ mesons have also been investigated. Nevertheless, a conclusive determination has not been achieved yet [24].

[*]Contact author: zhangyj@stu.cqu.edu.cn
[†]Contact author: wuxg@cqu.edu.cn
[‡]Contact author: hudd@stu.cqu.edu.cn
[§]Contact author: zlong@cqu.edu.cn

The quark-flavor (QF) scheme [25,26] and the singlet-octet (SO) scheme [27,28] are two commonly used schemes for describing the $\eta$ and $\eta'$ mixing. These two mixing schemes assume that the physical states $|\eta\rangle$ and $|\eta'\rangle$ are either a linear combination of $|\eta_8\rangle$- and $|\eta_1\rangle$- states for the SO scheme, or a linear combination of $|\eta_q\rangle$- and $|\eta_s\rangle$- states for the QF scheme. Two mixing schemes can be related via a proper rotation using an ideal mixing angle $\theta_i$, e.g., $\cos\theta_i = \sqrt{1/3}$ [29]. Till now, by taking more possible components into account, different mixings such as the $\eta-\eta'$ mixing [25,30,31], the $\eta-\eta'-\eta_c$ mixing [27,32–35] and the $\eta-\eta'-G$ mixing [12,36–39] and the $\eta-\eta'-G-\eta_c$ mixing [35] have been studied. Practically, the $B/D/D_s$ mesons' semileptonic decays $B\to\eta^{(\prime)}\ell^+\nu_\ell$ [40,41], $D\to\eta^{(\prime)}\ell^+\nu_\ell$ [42–45] and $D_s\to\eta^{(\prime)}\ell^+\nu_\ell$ [12,46,47] can serve as helpful platforms for exploring the differences among various mechanisms. Based on the mixing states, e.g., using $\phi$ to represent the mixing angle between $|\eta_q\rangle$ and $|\eta_s\rangle$ states, the heavy-to-light transition form factors (TFFs) $f^+_{B/D/D_s\to\eta^{(\prime)}}$ for the semileptonic decays $B/D/D_s\to\eta^{(\prime)}\ell^+\nu_\ell$ within the framework of the QF scheme satisfy the relations [48,49]

$$f^+_{B/D/D_s\to\eta} = f^+_{B/D/D_s\to\eta_q}\cos\phi - f^+_{B/D/D_s\to\eta_s}\sin\phi,$$
$$f^+_{B/D/D_s\to\eta'} = f^+_{B/D/D_s\to\eta_q}\sin\phi + f^+_{B/D/D_s\to\eta_s}\cos\phi.$$

The QCD light-cone sum rules (LCSR) method [50–53] offers an effective framework to calculate those heavy-to-light TFFs, e.g., the $B^+/D^+/D_s^+\to\eta_{s,q}$ TFFs, which determines the nonperturbative parameters of hadronic states and is applicable in both small and intermediate $q^2$-region, where $q^2$ stands for the momentum transfer between the heavy and the light mesons. Under the LCSR approach, by using the operator product expansion (OPE) near the light-cone $x^2\approx 0$, the nonperturbative hadronic matrix elements can be parametrized as the light-cone distribution amplitudes (LCDAs) of the light meson with various twist structures. Moreover, the light-meson LCDA can be calculated by using the QCD sum rules within the framework of background field theory (BFTSR) [54]. The BFTSR method provides a systematic description of vacuum condensates from a field-theoretic perspective and a viable approach to accounting for nonperturbative effects [55–59]. It assumes that quark and gluon fields consist of background fields and quantum fluctuations around them. By taking the QCD background field as the starting point for QCD sum rules, this method not only offers a clear physical picture but also greatly simplifies calculations—owing to its ability to impose distinct gauge conditions on quantum fluctuations and background fields, respectively. More specifically, the vacuum expectation values of background fields well describe nonperturbative effects, whereas quantum fluctuations represent calculable perturbative effects. In this paper, we will adopt the BFTSR to calculate the moments of $\eta_s$-meson's twist-2 LCDA and obtain the decay constant of $\eta_s$.

The remaining parts of this paper are organized as follows. In Sec. II, we give the calculation technology, including the $\eta-\eta'-G-\eta_c$ mixing formalism based on the quark-flavor base, a brief introduction of the LCSRs for the TFFs of the semileptonic decays $B^+/D^+/D_s^+\to\eta^{(\prime)}\ell^+\nu_\ell$, and the LCDAs for both $\eta_{s,q}$ and the gluonium state. In Sec. III, we give our numerical results for the mixing of $\eta-\eta'-G-\eta_c$, the $B^+/D^+/D_s^+\to\eta_{s,q}$ TFFs, and the decay widths/branching fractions of the semileptonic decays $B^+/D^+/D_s^+\to\eta^{(\prime)}\ell^+\nu_\ell$. Section IV is reserved for a summary.

## II. CALCULATION TECHNOLOGY

### A. $\eta$ and $\eta'$ mesons' mixing matrix

The conventional SO scheme and the QF scheme are two widely accepted schemes to study the $\eta$ and $\eta'$ mixing [60–62]. The SO scheme defines two hypothetical pure singlet and octet states, labeled by $|\eta_1\rangle = \frac{|\bar{u}u+\bar{d}d+\bar{s}s\rangle}{\sqrt{3}}$ and $|\eta_8\rangle = \frac{|\bar{u}u+\bar{d}d-2\bar{s}s\rangle}{\sqrt{6}}$, to describe the mixing of four decay constants with two mixing angles

$$\begin{pmatrix} f^8_\eta & f^1_\eta \\ f^8_{\eta'} & f^1_{\eta'} \end{pmatrix} = \begin{pmatrix} \cos\theta_8 & -\sin\theta_1 \\ \sin\theta_8 & \cos\theta_1 \end{pmatrix} \begin{pmatrix} f_8 & 0 \\ 0 & f_1 \end{pmatrix}, \qquad (1)$$

with $\langle 0|J^j_{\mu5}(0)|\eta_{8(1)}(p)\rangle = if_{\eta_{8(1)}}p^\mu$ and $\langle 0|J^j_{\mu5}(0)|\eta^{(\prime)}(p)\rangle = if^j_{\eta^{(\prime)}}p^\mu$, where $j=8,\ 1$ and $J^8_{\mu5} = \frac{(\bar{u}\gamma^\mu\gamma_5 u+\bar{d}\gamma^\mu\gamma_5 d-2\bar{s}\gamma^\mu\gamma_5 s)}{\sqrt{6}}$, $J^1_{\mu5} = \frac{(\bar{u}\gamma^\mu\gamma_5 u+\bar{d}\gamma^\mu\gamma_5 d+\bar{s}\gamma^\mu\gamma_5 s)}{\sqrt{3}}$. In the QF scheme, the physical meson states $|\eta\rangle$ and $|\eta'\rangle$ can be expressed as the linear combination of orthogonal states $|\eta_q\rangle = \frac{|\bar{u}u+\bar{d}d\rangle}{\sqrt{2}}$ and $|\eta_s\rangle = |\bar{s}s\rangle$ via the one-parameter matrix [27]

$$\begin{pmatrix} |\eta\rangle \\ |\eta'\rangle \end{pmatrix} = \begin{pmatrix} \cos\phi & -\sin\phi \\ \sin\phi & \cos\phi \end{pmatrix} \begin{pmatrix} |\eta_q\rangle \\ |\eta_s\rangle \end{pmatrix}. \qquad (2)$$

Considering the fact that, in the QF scheme the mixing of $\eta$ and $\eta'$ mesons is not governed by the $SU(3)_F$ breaking effects but by the Okubo-Zweig-Iizuka (OZI)-rule violating which has been proven to be small [27]. And the decay constants simply follow the same mixing form of particles with only one mixing angle. So for convenience, we will use the QF scheme in the following discussion.

There may have intrinsic charm and the gluonic components in the physical states $|\eta\rangle$ and $|\eta'\rangle$, we then extend the above matrix (2) to include the physical pseudoscalar glueball $|G\rangle$ and the pseudoscalar meson $|\eta_c\rangle$ components. For the possible mixing of $\eta - \eta' - \eta_c$, $\eta - \eta' - G$, and $\eta - \eta' - G - \eta_c$, the physical states $|\eta\rangle$, $|\eta'\rangle$, $|G\rangle$, and $|\eta_c\rangle$ are related to the octet state $|\eta_8\rangle$, the singlet state $|\eta_1\rangle$, the unmixed glueball state $|gg\rangle$. and the intrinsic charm component $|\eta_{c0}\rangle$ via the following rotations,

$$\begin{pmatrix} |\eta\rangle \\ |\eta'\rangle \\ |\eta_c\rangle \end{pmatrix} = U_3(\theta) U_1(\theta_c) \begin{pmatrix} |\eta_8\rangle \\ |\eta_1\rangle \\ |\eta_{c0}\rangle \end{pmatrix}, \tag{3}$$

$$\begin{pmatrix} |\eta\rangle \\ |\eta'\rangle \\ |G\rangle \end{pmatrix} = U_3(\theta) U_1(\phi_G) \begin{pmatrix} |\eta_8\rangle \\ |\eta_1\rangle \\ |gg\rangle \end{pmatrix}, \tag{4}$$

$$\begin{pmatrix} |\eta\rangle \\ |\eta'\rangle \\ |G\rangle \\ |\eta_c\rangle \end{pmatrix} = U_{34}(\theta) U_{14}(\phi_g) U_{13}(\phi_c) U_{12}(\theta_g) \begin{pmatrix} |\eta_8\rangle \\ |\eta_1\rangle \\ |gg\rangle \\ |\eta_{c0}\rangle \end{pmatrix}. \tag{5}$$

The transformation matrices $U_{1,3}$, which represent the rotations around the axis along the octet state $|\eta_8\rangle$, are defined as [63]

$$U_3(\theta) = \begin{pmatrix} \cos\theta & -\sin\theta & 0 \\ \sin\theta & \cos\theta & 0 \\ 0 & 0 & 1 \end{pmatrix}, \tag{6}$$

$$U_1(\phi_G) = \begin{pmatrix} 1 & 0 & 0 \\ 0 & \cos\phi_G & \sin\phi_G \\ 0 & -\sin\phi_G & \cos\phi_G \end{pmatrix}, \tag{7}$$

which are based on the assumption that $|\eta_8\rangle$ does not mix with the intrinsic charm component $|\eta_{c0}\rangle$ or the unmixed glueball state $|gg\rangle$. As for the four-particle mixing (5), the rotational matrices are constructed as follows:

$$U_{34}(\theta) = \begin{pmatrix} \cos\theta & -\sin\theta & 0 & 0 \\ \sin\theta & \cos\theta & 0 & 0 \\ 0 & 0 & 1 & 0 \\ 0 & 0 & 0 & 1 \end{pmatrix}, \tag{8}$$

$$U_{14}(\phi_g) = \begin{pmatrix} 1 & 0 & 0 & 0 \\ 0 & \cos\phi_g & \sin\phi_g & 0 \\ 0 & -\sin\phi_g & \cos\phi_g & 0 \\ 0 & 0 & 0 & 1 \end{pmatrix}, \tag{9}$$

$$U_{13}(\phi_c) = \begin{pmatrix} 1 & 0 & 0 & 0 \\ 0 & \cos\phi_c & 0 & \sin\phi_c \\ 0 & 0 & 1 & 0 \\ 0 & -\sin\phi_c & 0 & \cos\phi_c \end{pmatrix}, \tag{10}$$

$$U_{12}(\theta_g) = \begin{pmatrix} 1 & 0 & 0 & 0 \\ 0 & 1 & 0 & 0 \\ 0 & 0 & \cos\theta_g & \sin\theta_g \\ 0 & 0 & -\sin\theta_g & \cos\theta_g \end{pmatrix}, \tag{11}$$

where $\theta$, $\phi_g$, $\theta_c$, and $\theta_g$ are mixing angles among different components, which are to be determined. Because the flavor quantum numbers of $|gg\rangle$, $|\eta_{c0}\rangle$, $|\eta_1\rangle$, and $|\eta_8\rangle$ are not identical. Specifically, the unmixed pseudoscalar glueball state $|gg\rangle$, $|\eta_1\rangle$, and $|\eta_{c0}\rangle$ are both SU(3) flavor singlets, while $|\eta_8\rangle$ belongs to an SU(3) flavor octet. So, here we have implicitly adopted the same assumption as that of Ref. [63] to construct these matrix elements, i.e., the octet state $|\eta_8\rangle$ does not mix with the glueball state $|gg\rangle$ and the intrinsic charm state $|\eta_{c0}\rangle$. However, there is a slight difference from the transition matrix in the Ref. [35]: we have retain the mixing between $|\eta_{c0}\rangle$ and $|\eta_1\rangle$, whereas the reference assumes that heavy quark state $|\eta_{c0}\rangle$ mainly mix with the unmixed glueball state and neglects the mixing between heavy quark state and the $|\eta_1\rangle$ state.

Additionally, the octet and singlet states have the following transform relation with the flavor states:

$$\begin{pmatrix} |\eta_8\rangle \\ |\eta_1\rangle \\ |gg\rangle \\ |\eta_{c0}\rangle \end{pmatrix} = U_{34}(\theta_i) \begin{pmatrix} |\eta_q\rangle \\ |\eta_s\rangle \\ |gg\rangle \\ |\eta_{c0}\rangle \end{pmatrix}, \tag{12}$$

where $\theta_i = 54.7°$ is the ideal mixing angle. Using the transformation (12), we then obtain the following transformation matrix for the flavor states to be transformed into the physical states

$$
\begin{aligned}
&U(\theta,\phi_g,\phi_c,\theta_g)\\
&=U_{34}(\theta)U_{14}(\phi_g)U_{13}(\phi_c)U_{12}(\theta_g)U_{34}(\theta_i)\\
&=\begin{pmatrix}
c\theta c\theta_i - s\theta s\theta_i c\phi_c c\phi_g & -s\theta c\theta_i c\phi_c c\phi_g - c\theta s\theta_i & s\theta(s\theta_g s\phi_c c\phi_g - c\theta_g s\phi_g) & -s\theta(c\theta_g s\phi_c c\phi_g + s\theta_g s\phi_g)\\
c\theta s\theta_i c\phi_c c\phi_g + s\theta c\theta_i & c\theta c\theta_i c\phi_c c\phi_g - s\theta s\theta_i & c\theta(c\theta_g s\phi_g - s\theta_g s\phi_c)c\phi_g & c\theta(c\theta_g s\phi_c c\phi_g + s\theta_g s\phi/_g)\\
-s\theta_i c\phi_c s\phi_g & -c\theta_i c\phi_c s\phi_g & s\theta_g s\phi_c s\phi_g + c\theta_g c\phi_g & s\theta_g c\phi_g - c\theta_g s\phi_c s\phi_g\\
-s\theta_i s\phi_c & -c\theta_i s\phi_c & -s\theta_g c\phi_c & c\theta_g c\phi_c
\end{pmatrix},
\end{aligned}
\tag{13}
$$

where for convenience, we have adopted the short notations, $c \equiv \cos(\cdot)$ and $s \equiv \sin(\cdot)$.

The required decay constants $f_q$, $f_s$, and $f_c$ are defined by the following matrix elements [64]

$$
\begin{aligned}
\langle 0|\frac{(\bar{u}\gamma^\mu\gamma_5 u + \bar{d}\gamma^\mu\gamma_5 d)}{\sqrt{2}}|\eta_q(p)\rangle &= i f_q p^\mu,\\
\langle 0|\bar{s}\gamma^\mu\gamma_5 s|\eta_s(p)\rangle &= i f_s p^\mu,\\
\langle 0|\bar{c}\gamma^\mu\gamma_5 c|\eta_{c0}(p)\rangle &= i f_c p^\mu.
\end{aligned}
\tag{14}
$$

where for short, we have used $f_q = f_{\eta_q}$, $f_s = f_{\eta_s}$, and $f_c = f_{\eta_{c0}}$. The derivatives of the axial vector currents between vacuum state and particle states are

$$
\begin{aligned}
\partial_\mu(\bar{q}\gamma^\mu\gamma_5 q) &= 2im_q\bar{q}\gamma_5 q + \frac{\alpha_s}{4\pi}G_{\mu\nu}\tilde{G}^{\mu\nu},\\
\partial_\mu(\bar{s}\gamma^\mu\gamma_5 s) &= 2im_s\bar{s}\gamma_5 s + \frac{\alpha_s}{4\pi}G_{\mu\nu}\tilde{G}^{\mu\nu},\\
\partial_\mu(\bar{c}\gamma^\mu\gamma_5 c) &= 2im_c\bar{c}\gamma_5 c + \frac{\alpha_s}{4\pi}G_{\mu\nu}\tilde{G}^{\mu\nu},
\end{aligned}
\tag{15}
$$

where $m_{q,s,c}$ are current quark masses, $G_{\mu\nu}$ is the field-strength tensor and $\tilde{G}^{\mu\nu}$ is the dual field-strength tensor. The vacuum-to-meson transition matrix elements of the derivatives of axial vector current are given by the product of the squared meson mass $m_P^2$ and the decay constant as $\langle 0|\partial_\mu J_{\mu 5}^{q(s,c)}|P(p)\rangle = m_P^2 f_P^{q(s,c)}$.

Following the procedures in Ref. [65], we get the mass matrix for $\eta - \eta' - G - \eta_c$ mixing:

$$
M_{qsgc} = U^\dagger(\theta,\phi_g,\phi_c,\theta_g)M^2 U(\theta,\phi_g,\phi_c,\theta_g)\tilde{J},
\tag{16}
$$

where the matrices are

$$
M_{qsgc} = \begin{pmatrix}
m_{qq}^2 + \sqrt{2}G_q/f_q & m_{sq}^2 + G_q/f_s & m_{cq}^2 + G_q/f_c\\
m_{qs}^2 + \sqrt{2}G_s/f_q & m_{ss}^2 + G_s/f_s & m_{cs}^2 + G_s/f_c\\
m_{qg}^2 + \sqrt{2}G_g/f_q & m_{sg}^2 + G_g/f_s & m_{cg}^2 + G_g/f_c\\
m_{qc}^2 + \sqrt{2}G_c/f_q & m_{sc}^2 + G_c/f_s & m_{cc}^2 + G_c/f_c
\end{pmatrix},
\tag{17}
$$

$$
M^2 = \begin{pmatrix}
m_\eta^2 & 0 & 0 & 0\\
0 & m_{\eta'}^2 & 0 & 0\\
0 & 0 & m_G^2 & 0\\
0 & 0 & 0 & m_{\eta_c}^2
\end{pmatrix},
\tag{18}
$$

$$
\tilde{J} = \begin{pmatrix}
1 & f_q^s/f_s & f_q^c/f_c\\
f_s^q/f_q & 1 & f_s^c/f_c\\
f_g^q/f_q & f_g^s/f_s & f_g^c/f_c\\
f_c^q/f_q & f_c^s/f_s & 1
\end{pmatrix}.
\tag{19}
$$

We will take the values of the decay constants $f_{s,g,c}^q$, $f_{q,g,c}^s$, and $f_{q,s,g}^c$ as zero in calculation, since they are suppressed by the QZI rule. The abbreviations $m_{qq,\ldots}^2$ and $G_{q,s,g,c}$ stand for the pseudoscalar densities and the $U(1)$ anomaly matrix elements, respectively, which are defined as the following form:

$$
\begin{aligned}
m_{qq,qs,qg,qc}^2 &\equiv \frac{\sqrt{2}}{f_q}\langle 0|m_u\bar{u}i\gamma_5 u + m_d\bar{d}i\gamma_5 d|\eta_q,\eta_s,g,\eta_c\rangle,\\
m_{sq,ss,sg,sc}^2 &\equiv \frac{2}{f_s}\langle 0|m_s\bar{s}i\gamma_5 s|\eta_q,\eta_s,g,\eta_c\rangle,\\
m_{cq,cs,cg,cc}^2 &\equiv \frac{2}{f_c}\langle 0|m_c\bar{c}i\gamma_5 c|\eta_q,\eta_s,g,\eta_c\rangle,\\
G_{q,s,g,c} &\equiv \langle 0|\alpha_s G\tilde{G}/(4\pi)|\eta_q,\eta_s,g,\eta_c\rangle.
\end{aligned}
\tag{20}
$$

For convenience, we put the expressions for the required mass matrix elements that are related to the mixing angles in the Appendix. Numerically, we have observed that the mixing of intrinsic charm component with $\eta_{q,s}$ are much larger than the case of gluonic component with $\eta_{q,s}$, but the charm component does not have contribution to the TFFs, so we will concentrate on the gluonic component's effect and will not discuss the mixing effect from the intrinsic charm component.

### B. The $H \to \eta^{(\prime)}$ TFFs using the LCSR

In this subsection, we give a brief introduction of our calculation technology for $H \to \eta^{(\prime)}$ TFFs by using the LCSR approach. Here "$H$" symbolizes the $B^+$, $D^+$, and $D_s^+$, respectively. The $H \to \eta^{(\prime)}$ TFFs can be defined by the local $H \to \eta^{(\prime)}$ matrix element [38]

$$\langle\eta^{(\prime)}|V_\mu^{\eta^{(\prime)}}|H(p+q)\rangle = 2p_\mu f^+_{H\to\eta^{(\prime)}}(q^2) + q_\mu(f^+_{H\to\eta^{(\prime)}}(q^2) + f^-_{H\to\eta^{(\prime)}}(q^2)), \tag{21}$$

where $V_\mu^{\eta^{(\prime)}}$ is the interaction vertex, which is listed in Table I. To derive their LCSR expressions, we construct the following correlation function (correlator):

$$\Pi_\mu(p,q) = i\int d^4x e^{iq\cdot x}\langle\eta^{(\prime)}|T\{V_\mu^{\eta^{(\prime)}}(x), j_H^\dagger(0)\}|0\rangle = \Pi[q^2,(p+q)^2]p_\mu + \tilde{\Pi}[q^2,(p+q)^2]q_\mu. \tag{22}$$

where $j_H$ is the local interpolating current [49], which is also listed in Table I.

TABLE I. The currents $J_H$ and the interaction vertex $V_\mu^{\eta^{(\prime)}}$, where $H$ represents $B^+$, $D^+$, and $D_s^+$, respectively.

| Decay | $j_H$ | Interaction vertex |
|---|---|---|
| $B^+ \to \eta^{(\prime)}\ell^+\nu_\ell$ | $j_{B^+} = m_b\bar{u}i\gamma_5 b$ | $V_\mu^{\eta^{(\prime)}} = \bar{u}(x)\gamma_\mu b(x)$ |
| $D^+ \to \eta^{(\prime)}\ell^+\nu_\ell$ | $j_{D^+} = m_c\bar{d}i\gamma_5 c$ | $V_\mu^{\eta^{(\prime)}} = \bar{d}(x)\gamma_\mu c(x)$ |
| $D_s^+ \to \eta^{(\prime)}\ell^+\nu_\ell$ | $j_{D_s^+} = m_c\bar{s}i\gamma_5 c$ | $V_\mu^{\eta^{(\prime)}} = \bar{s}(x)\gamma_\mu c(x)$ |

To deal with the correlator (22) in the timelike region, we can insert a complete set of the intermediate hadronic states in the correlator. Isolating the pole term (corresponding to the ground state), we obtain its hadronic representation by integrating over the hadronic spectral density. The hadronic expression can be written as

$$\Pi[q^2,(p+q)^2] = \frac{2m_H^2 f_H f^+_{H\to\eta^{(\prime)}}(q^2)}{[m_H^2-(p+q)^2]}p_\mu + \int_{s_0}^\infty ds\frac{\rho^{\mathcal{H}}(q^2,s)}{s-(p+q)^2}, \tag{23}$$

where $s_0$ is the continuum threshold parameter and $\rho^{\mathcal{H}}$ is the hadronic spectral density.

In the spacelike region, one can adopt OPE to calculate the correlator, which results in a convolution of the perturbatively calculable hard-scattering amplitude and the universal soft LCDAs near the light cone. And the light-cone expansion result for the correlation function can be written as

$$\Pi^{\rm OPE}[q^2,(p+q)^2] = F_0(q^2,(p+q)^2) + \frac{\alpha_s C_F}{4\pi}(F_1(q^2,(p+q)^2) + F_1^{gg,+}(q^2,(p+q)^2)), \tag{24}$$

where the first term is the leading-order (LO) contributions for all the LCDAs and the second term is the next-to-leading order (NLO) contribution. The third term represent the gluonic contribution, which enters at the NLO level [39].

Finally, by applying the Borel transformation and subtracting the contributions from higher resonances and continuum states, we can get the LCSR for the TFFs

$$f^+_{H\to\eta^{(\prime)}}(q^2) = \frac{e^{m_H^2/M^2}}{2m_H^2 f_H}\bigg[F_0(q^2,M^2,s_0) + \frac{\alpha_s C_F}{4\pi}(F_1(q^2,M^2,s_0) + F_1^{gg,+}(q^2,M^2,s_0))\bigg]. \tag{25}$$

In this equation, the NLO amplitude $F_1(q^2,M^2,s_0)$ can be found in Ref. [66], which is given as a factorized form of the convolutions. And the expression for the gluonic contribution has been presented in Ref. [39], i.e.,

$$F_1^{gg,+}(q^2,M^2,s_0) = f^1_{\eta^{(\prime)}}a^{\eta^{(\prime)}}_{2;g}\frac{1}{C_F}\int_{m^2}^{s_0^M} e^{-s/M^2}f^{gg,+}(s,q^2), \tag{26}$$

where the Gegenbauer moments $a^{\eta^{(\prime)}}_{2;g}$ will be defined in the next subsection, and

$$f^{gg,+} = 20m^2\frac{s-m^2}{27\sqrt{3}(s-q^2)^5}\bigg\{3(m^2-q^2)(5m^4-5m^2(q^2+s)+q^4+3q^2s+s^2)\bigg[2\ln\bigg(\frac{s-m^2}{m^2}\bigg)-\ln\bigg(\frac{\mu^2}{m^2}\bigg)\bigg] - (37m^6-m^4(56q^2+55s)+m^2(18q^4+76q^2s+17s^2)+3q^6-27q^4s-11q^2s^2-2s^3)\bigg\}, \tag{27}$$

where $m = m_b$ or $m_c$ for $B^+$, $D^+$, and $D_s^+$ mesons, respectively.

In addition, based on the definition of the TFFs, we can deduce that the TFFs can be expressed in the following mixing form

$$\begin{pmatrix} f^+_{H\to\eta} \\ f^+_{H\to\eta'} \end{pmatrix} = U(\phi) \begin{pmatrix} f^+_{H\to\eta_q} \\ f^+_{H\to\eta_s} \end{pmatrix}, \tag{28}$$

with $f^+_{H\to\eta_q} = f^{(\bar{q}q),+}_{H\to\eta_q} + f^{(gg),+}_{H\to\eta_q}$, $f^+_{H\to\eta_s} = f^{(\bar{s}s),+}_{H\to\eta_s} + f^{(gg),+}_{H\to\eta_s}$. The LCSRs for the $H\to\eta_{q(s)}$ TFFs have been derived in our previous works, e.g., Refs. [66,67].

Compared with the leading-twist terms, the high-twist terms will be power suppressed and have small contributions. So we will only consider the twist-3 and twist-4 LCDAs, whose expressions and input parameters are taken from Ref. [39].

Using the $B^+/D^+/D_s^+\to\eta^{(\prime)}$ TFFs, we derive the decay widths and the decay branching fractions via the following equations [68]

$$\begin{aligned} &\frac{d\Gamma}{dq^2}(H\to\eta^{(\prime)}\ell^+\nu_\ell) \\ &= \frac{G_F^2|V_{ub(cd;cs)}|^2}{192\pi^3 m_H^3}\int_{m_\ell^2}^{(m_H-m^2_{\eta^{(\prime)}})^2} \\ &\quad\times[(m_H^2+m^2_{\eta^{(\prime)}}-q^2)^2-4m_H^2 m^2_{\eta'}]^{3/2}|f^+_{H\to\eta^{(\prime)}}(q^2)|^2 \end{aligned} \tag{29}$$

and

$$\frac{\mathcal{B}(H\to\eta^{(\prime)}\ell^+\nu_\ell)}{\tau(H)} = \int_0^{q^2_{\max}}\frac{d\Gamma}{dq^2}(H\to\eta^{(\prime)}\ell^+\nu_\ell), \tag{30}$$

where $q^2_{\max} = (m_H - m_{\eta^{(\prime)}})^2$, and $\tau(H)$ represents the lifetime of $H$-meson.

### C. Moments of the $\eta_{q,s}$-meson leading-twist LCDAs and their scale-running behaviors

With the QCD theory in background field, we can derive the moments of $\eta_{q,s}$-mesons' leading-twist LCDAs and obtain their decay constants. In this subsection, we give the main results for the derivation of the moments of $\eta_s$-meson leading-twist LCDA. The results of $\eta_q$ can be derived via the same way, which have been done in Ref. [67]. And if not specially stated, we will direct adopt the $\eta_q$ results of Ref. [67] to do our numerical calculations throughout the paper.

Following the standard BFTSR procedures, we take its correlation function (correlator) as

$$\begin{aligned} \Pi_{(n,0);\eta_s}(z,q) &= i\int d^4x e^{iq\cdot x}\langle 0|T\{J_n(x),J_0^\dagger(0)\}|0\rangle \\ &= (z\cdot q)^{n+2}\Pi_{(n,0);\eta_s}(q^2), \end{aligned} \tag{31}$$

where the currents $J_n(x) = \bar{s}(x)\mathcal{C}_s \not{z}\gamma_5(iz\cdot\overleftrightarrow{D})^n s(x)$ and $J_0^\dagger(0) = \bar{s}(0)\mathcal{C}_s\not{z}\gamma_5 s(0)$ with $z^2=0$. And $\mathcal{C}_s = (\mathcal{C}_1 - \sqrt{2}\mathcal{C}_8)/\sqrt{3}$ with $\mathcal{C}_1 = \mathbf{1}/\sqrt{3}$ and $\mathcal{C}_8 = \lambda_8/\sqrt{2}$ where $\lambda_8$ is the Gell-Mann matrix and 1 is the $3\times 3$ unit matrix. The specific expression of $(iz\cdot\overleftrightarrow{D})^n = (iz\cdot\overrightarrow{D} - iz\cdot\overleftarrow{D})^n$ can be found in Ref. [69]. Due to the conservation of the $G$-parity, only even moments are nonzero, which indicate $n = 0, 2, 4, \ldots$. The operator product expansion (OPE) can be used to deal with the correlator (31) in the deep Euclidean region $q^2 \ll 0$. And the result can be expressed as a expansion series over the basic vacuum condensates with increasing dimensions, whose explicit expression have been given in Refs. [66].

On the other hand, the correlator (31) can also be calculated by inserting a complete set of the intermediate hadronic states in the physical region to get the hadron expression. By using the conventional quark-hadron duality [59], the hadronic expression of the correlator can be written as

$$\begin{aligned} \mathrm{Im}\, I^{\mathrm{Had}}_{(n,0);\eta_s}(q^2) &= \pi\delta(q^2-m^2_{\eta_s})f^2_{\eta_s}\langle\xi_{n;\eta_s}\rangle\langle\xi_{0;\eta_s}\rangle \\ &\quad+\pi\frac{3}{4\pi^2(n+1)(n+3)}\theta(q^2-s_{\eta_s}), \end{aligned} \tag{32}$$

where $f_{\eta_s}$ is the decay constant and $s_{\eta_s}$ is the continuum threshold for the lowest continuum state. Here we have used the definition

$$\langle\xi_{n;\eta_s}\rangle = \int_0^1 dx\xi^n\phi_{2;\eta_s}(x), \tag{33}$$

where $\phi_{2;\eta_s}(x)$ is the leading-twist LCDA of $\eta_s$ and $\xi = 2x-1$. Using the dispersion relation

$$\frac{1}{\pi}\int_{4m_s^2}^{\infty}ds\frac{\mathrm{Im}\, I^{\mathrm{Had}}_{(n,0);\eta_s}(s)}{s-q^2} = I^{\mathrm{OPE}}_{(n,0);\eta_s}(q^2), \tag{34}$$

we then obtain the expression of $\langle\xi_{n;\eta_s}\rangle$ by matching the hadronic expression with the OPE result. And by further applying the Borel transformation, the uncertainties caused by the unwanted contributions from the higher-order dimensional vacuum condensates and the continuum states can be further suppressed. And the resultant sum rule for $\langle\xi_{n;\eta_s}\rangle$ becomes

$$
\begin{aligned}
&\frac{f_{\eta_s}^2 \langle \xi_{n;\eta_s} \rangle \langle \xi_{0;\eta_s} \rangle}{M^2 e^{m_{\eta_s}^2/M^2}} \\
&= \frac{1}{\pi} \frac{1}{M^2} \int_{4m_s^2}^{s_{\eta_s}} ds e^{-s/M^2} \frac{3 v^{n+1}}{8\pi (n+1)(n+3)} \left(1 + \frac{\alpha_s}{\pi} A_n'\right) \left\{ [1 + (-1)^n](n+1) \frac{1 - v^2}{2} + [1 + (-1)^n] \right\} \\
&\quad + \frac{2 m_s \langle \bar{s} s \rangle}{M^4} + \frac{\langle \alpha_s G^2 \rangle}{12\pi M^4} \frac{1 + n\theta(n-2)}{n+1} - \frac{8n+1}{9} \frac{m_s \langle g_s \bar{s} \sigma T G s \rangle}{M^6} + \frac{\langle g_s \bar{s} s \rangle}{81 M^6} 4(2n+1) - \frac{\langle g_s^3 f G^3 \rangle}{48\pi^2 M^6} n\theta(n-2) \\
&\quad + \frac{\sum \langle g_s^2 \bar{s} s \rangle^2}{486\pi^2 M^6} \left\{ -2(51n+25)\left(-\ln \frac{M^2}{\mu^2}\right) + 3(17n+35) + \theta(n-2) \left[ 2n \left(-\ln \frac{M^2}{\mu^2}\right) - 25(2n+1)\tilde{\psi}(n) \right.\right. \\
&\quad \left.\left. + \frac{1}{n}(49n^2 + 100n + 56) \right] \right\} + m_s^2 \left\{ -\frac{\langle \alpha_s G^2 \rangle}{6\pi M^6} \left[ \theta(n-2)(n\tilde{\psi}(n) - 2) + 2n \left(-\ln \frac{M^2}{\mu^2}\right) - n - 2 \right] + \frac{\langle g_s^3 f G^3 \rangle}{288\pi^2 M^8} \right. \\
&\quad \times \left\{ -10\delta^{n0} + \theta(n-2) \left[ 4n(2n-1)\left(-\ln \frac{M^2}{\mu^2}\right) - 4n\tilde{\psi}(n) + 8(n^2 - n + 1) \right] + \theta(n-4)[2n(8n-1)\tilde{\psi}(n) \right. \\
&\quad \left. - (19n^2 + 19n + 6)] + 8n(3n-1)\left(-\ln \frac{M^2}{\mu^2}\right) - (21n^2 + 53n - 6) \right\} - \frac{\sum \langle g_s^2 s \bar{s} \rangle^2}{972\pi^2 M^8} \left\{ 6\delta^{n0} \left[ 16 \left(-\ln \frac{M^2}{\mu^2}\right) - 3 \right] \right. \\
&\quad + \theta(n-2) \left[ 8(n^2 + 12n - 12)\left(-\ln \frac{M^2}{\mu^2}\right) - 2(29n+22)\tilde{\psi}(n) + 4\left(5n^2 - 2n - 33 + \frac{46}{n}\right) \right] + \theta(n-4) \\
&\quad \left. \times [2(56n^2 - 25n + 24)\tilde{\psi}(n)(139n^2 + 91n + 54)] + 8(27n^2 - 15n - 11)\left(-\ln \frac{M^2}{\mu^2}\right) - 3(63n^2 + 159n - 50) \right\} \\
&\quad \left. + \frac{4(n-1)}{3} \frac{m_s \langle \bar{s} s \rangle}{M^6} + \frac{8n-3}{9} \frac{m_s \langle g_s \bar{s} \sigma T G s \rangle}{M^8} - \frac{4(2n+1)}{81} \frac{\langle g_s \bar{s} s \rangle^2}{M^8} \right\},
\end{aligned}
\tag{35}
$$

where

$$
\tilde{\psi}(n) = \psi\left(\frac{n+1}{2}\right) - \psi\left(\frac{n}{2}\right) + \ln 4. \tag{36}
$$

Here $v^2 = 1–4m_s^2/s$ and $A_n'$ represents the next-to-leading order radiative corrections to the perturbative part, whose first several values are $A_0' = 0$, $A_2' = 5/3$, $A_4' = 59/27$, $A_6' = 353/135$ [70], respectively. Moreover, it is found that the sum rule for the $\eta_s$ decay constant can be achieved by setting $n = 0$ to the above sum rule (35). Since the $0_{\rm th}$-order moment cannot be strictly normalized for fixed-order series, we suggest to use the ratio of the two sum rules, e.g., $\langle \xi_{n;\eta_s} \rangle = \langle \xi_{n;\eta_s} \rangle \langle \xi_{0;\eta_s} \rangle / \sqrt{(\langle \xi_{0;\eta_s} \rangle)^2}$ to do the calculation [71,72].

The leading-twist LCDAs of $\eta_q$ and $\eta_s$ are generally expanded as Gegenbauer polynomials [73]

$$
\phi_{2;\eta_{q(s)}}(x) = 6x\bar{x}\left[1 + \sum_{n=2,4,\ldots} a_{n;\eta_{q(s)}} C_n^{3/2}(x - \bar{x})\right], \tag{37}
$$

where $x$ and $\bar{x} = 1 - x$ are momentum fractions of light quark and antiquark inside $\eta_{q,(s)}$. $C_n^{3/2}$ and the following $C_n^{5/2}$ are Gegenbauer polynomials. The Gegenbauer moments $a_{n;\eta_{q(s)}}$ are scale dependent and their values at an arbitrary scale $\mu$ can be derived from their values at an initial scale $\mu_0$ by using the evolution equation, e.g., $a_{n;\eta_{q(s)}}(\mu) = (\alpha_s(\mu^2)/\alpha_s(\mu_0^2))^{\gamma_n/\beta_0} a_{n;\eta_{q(s)}}(\mu_0)$, where $\beta_0 = 11 - 2n_f/3$ for $SU_c(3)$ color group with $n_f$ active light flavors. Furthermore, the first three nonzero Gegenbauer moments $a_{n;\eta_{q(s)}}$ can be derived from the $\langle \xi_{n;\eta_{q(s)}} \rangle$-moments via the following relations:

$$
a_{2;\eta_{q(s)}} = \frac{35}{12}\langle \xi_{2;\eta_{q(s)}} \rangle - \frac{7}{12}, \tag{38}
$$

$$
a_{4;\eta_{q(s)}} = \frac{77}{8}\langle \xi_{4;\eta_{q(s)}} \rangle - \frac{77}{12}\langle \xi_{2;\eta_{q(s)}} \rangle + \frac{11}{24}, \tag{39}
$$

$$
\begin{aligned}
a_{6;\eta_{q(s)}} = {}& \frac{2145}{64}\langle \xi_{6;\eta_{q(s)}} \rangle + \frac{675}{64}\langle \xi_{2;\eta_{q(s)}} \rangle \\
& - \frac{2475}{64}\langle \xi_{4;\eta_{q(s)}} \rangle - \frac{25}{64}.
\end{aligned}
\tag{40}
$$

As for the leading-twist LCDA of the gluonic component $|gg\rangle$, we adopt the following forms suggested in Refs. [36,39,74,75]

$$
\phi_{2;g}(x) = x^2\bar{x}^2 \sum_{n=2,4,\ldots} a_{n;g} C_{n-1}^{5/2}(x - \bar{x}). \tag{41}
$$

The Gegenbauer moments $a_{n;\eta_{q(s)}}$ and $a_{n;g}$ are scale dependent. According to the analysis in Refs. [36,38], to make the OZI-violating effect small enough, $a_{n;\eta_8} \equiv a_{n;\eta_1}$ is required in the QF scheme, which implies $a_{n;\eta_{q(s)}}(\mu) = a_{n;\eta_1}(\mu)$. So at leading-order logarithmic accuracy, the gluonic component will affect the scale-running behavior of $a_{n;\eta_{q(s)}}$ via the following way [36]

$$a_{n;\eta_{q(s)}}(\mu) = a_{n;\eta_{q(s)}}(\mu_0)\left(\frac{\alpha_s(\mu_0)}{\alpha_s(\mu)}\right)^{\frac{\gamma_n^{(+)}}{\beta_0}} + \rho_n^{(-)} a_{n;g}(\mu_0)\left(\frac{\alpha_s(\mu_0)}{\alpha_s(\mu)}\right)^{\frac{\gamma_n^{(-)}}{\beta_0}},$$
$$a_{n;g}(\mu) = \rho_n^{(+)} a_{n;\eta_{q(s)}}(\mu_0)\left(\frac{\alpha_s(\mu_0)}{\alpha_s(\mu)}\right)^{\frac{\gamma_n^{(+)}}{\beta_0}} + a_{n;g}(\mu_0)\left(\frac{\alpha_s(\mu_0)}{\alpha_s(\mu)}\right)^{\frac{\gamma_n^{(-)}}{\beta_0}}, \tag{42}$$

where the parameters $\gamma_n^{(\pm)}$ and $\rho_n^{(\pm)}$ are

$$\gamma_n^{(\pm)} = \frac{1}{2}\left[\gamma_n^{qq} + \gamma_n^{gg} \pm \sqrt{(\gamma_n^{qq} - \gamma_n^{gg})^2 + 4\gamma_n^{gq}\gamma_n^{qg}}\right], \tag{43}$$

with

$$\gamma_n^{qq} = C_F\left[3 + \frac{2}{(n+1)(n+2)} - 4\sum_{i=1}^{n+1}\frac{1}{i}\right], \tag{44}$$

$$\gamma_n^{qg} = \sqrt{n_f C_F}\,\frac{n(n+3)}{3(n+1)(n+2)}\, n \geq 2, \tag{45}$$

$$\gamma_n^{gq} = \sqrt{n_f C_F}\,\frac{12}{(n+1)(n+2)}\, n \geq 2, \tag{46}$$

$$\gamma_n^{gg} = \beta_0 + N_c\left[\frac{8}{(n+1)(n+2)} - 4\sum_{i=1}^{n+1}\frac{1}{i}\right] n \geq 2, \tag{47}$$

and

$$\rho_n^{(+)} = 6\,\frac{\gamma_n^{gq}}{\gamma_n^{(+)} - \gamma_n^{gg}}, \tag{48}$$

$$\rho_n^{(-)} = \frac{1}{6}\,\frac{\gamma_n^{qg}}{\gamma_n^{(-)} - \gamma_n^{qq}}. \tag{49}$$

The expressions for the two-particle twist-3 LCDAs considered the $m^2_{\eta^{(\prime)}}$ corrections are

$$\phi^p_{3q(s)} = h_{q(s)} + 60 m_{q(s)} f_{3q(s)} C_2^{1/2}(2u-1), \tag{50}$$

$$\phi^\sigma_{3q(s)} = 6u(1-u)[h_{q(s)} + 10 m_{q(s)} f_{3q(s)} C_2^{3/2}(2u-1)], \tag{51}$$

where

$$h_q = f_q(m_\eta^2\cos^2\phi + m_{\eta'}^2\sin^2\phi) - \sqrt{2} f_s(m_{\eta'}^2 - m_\eta^2)\sin\phi\cos\phi, \tag{52}$$

$$h_s = f_s(m_{\eta'}^2\cos^2\phi + m_\eta^2\sin^2\phi) - \frac{f_q}{\sqrt{2}}(m_{\eta'}^2 - m_\eta^2)\sin\phi\cos\phi. \tag{53}$$

## III. NUMERICAL ANALYSIS

### A. Input parameters

The parameters used in the numerical calculation are as follows. According to the PDG [13], we take the charm-quark running mass $m_c(\bar{m}_c) = 1.273^{+0.0046}_{-0.0046}$ GeV, the $b$-quark running mass $m_b(\bar{m}_b) = 4.183^{+0.007}_{-0.007}$ GeV, and the $s$-quark running mass $m_s(\bar{m}_s) = 0.0935^{+0.0008}_{-0.0008}$ GeV; the $\eta$, $\eta'$, $B$, $D$, and $D_s$-meson masses are $m_\eta = 0.5479$ GeV, $m_{\eta'} = 0.9578$ GeV, $m_{B^\pm} = 5.2794$ GeV, $m_{D^\pm} = 1.8695$ GeV, and $m_{D_s^\pm} = 1.9684$ GeV, respectively; the lifetimes of $B^\pm$, $D^\pm$, and $D_s^\pm$ mesons are $\tau(B^\pm) = 1.638 \pm 0.004$ ps, $\tau(D^\pm) = 1.033 \pm 0.005$ ps, and $\tau(D_s^\pm) = 0.5012 \pm 0.0022$ ps, respectively; the current-quark-masses for the light $u$-quark and $d$-quark at the scale $\mu = 2$ GeV are $m_u = 2.16^{+0.07}_{-0.07}$ MeV and $m_d = 4.7^{+0.07}_{-0.07}$ MeV; and the CKM matrix elements are $|V_{\rm ub}| = (3.82 \pm 0.20) \times 10^{-3}$, $|V_{cd}| = 0.221 \pm 0.004$, and $|V_{\rm cs}| = 0.975 \pm 0.006$. As for the decay constants $f_B$, $f_D$, $f_{D_s}$ and $f_{\eta_q}$, we take $f_B = 0.215^{+0.007}_{-0.007}$ GeV [39], $f_D = 0.142 \pm 0.006$ GeV [68], $f_{D_s} = 0.274 \pm 0.013 \pm 0.007$ GeV [76], and $f_{\eta_q} = 0.141 \pm 0.005$ GeV [67]. The renormalization scale is set as the typical momentum flow $\mu_B = \sqrt{m_B^2 - \bar{m}_b^2} \approx 3.2$ GeV for $B$-meson decay, $\mu_D = \sqrt{m_D^2 - \bar{m}_c^2} \approx 1.4$ GeV for $D$-meson decay, and $\mu_{D_s} = \sqrt{m_{D_s}^2 - \bar{m}_c^2} \approx 1.5$ GeV for $D_s$-meson decay. The input parameters for the twist-3 LCDAs are taken as $f_{3q} \backsimeq f_{3\pi} = 0.0045$ and $f_{3s} \backsimeq f_{3K} = 0.0045$ at the scale $\mu = 1$ GeV [39]. The values of the nonperturbative vacuum condensates that appear in Eq. (35) can be found in Refs. [66,77–79]. Meanwhile, each vacuum condensates and current quark masses should be run from the initial scale ($\mu_0 = 1$ GeV) to the required scale by using the renormalization group equations [71].

### B. The $\eta_s$ decay constant and the moments $\langle\xi_{n;\eta_s}\rangle$

The sum rule of $\eta_s$ decay constant can be achieved by setting $n = 0$ to Eq. (35). Using this sum rule to fix the $f_{\eta_s}$, we set the continuum threshold $s_0 = 1.5 \pm 0.1$ GeV$^2$ [80].

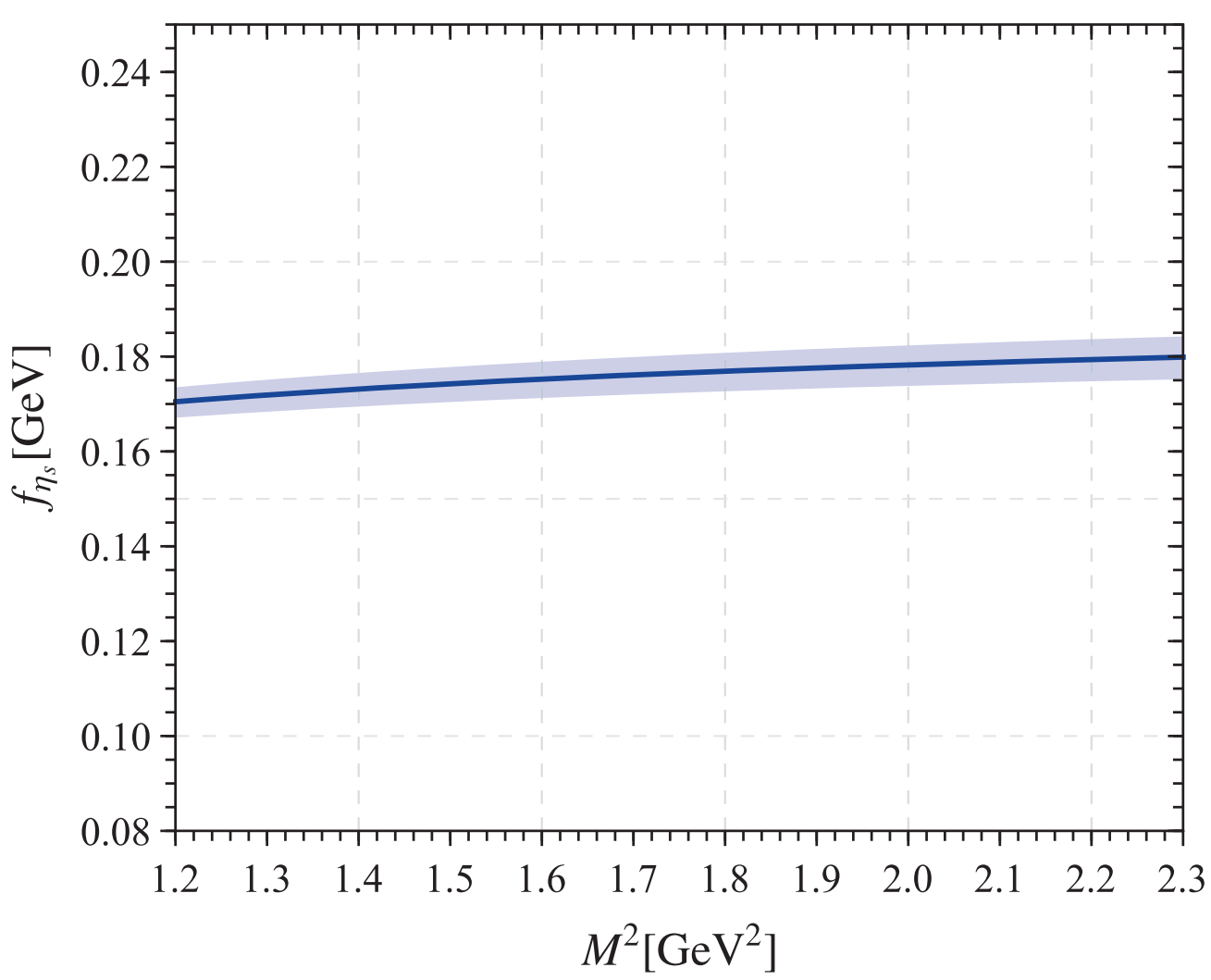

FIG. 1. The decay constant $f_{\eta_s}$ versus with the Borel parameter $M^2$, where the shaded band indicates the total uncertainties from all the mentioned input parameters.

To find the allowable window for the Borel parameter $M^2$, we adopt the following criteria,

(i) The continuum contribution is less than 30%;
(ii) The contribution of the dimension-six condensates is no higher than 5%;
(iii) The value of $f_{\eta_s}$ is stable in the Borel window.

The decay constant $f_{\eta_s}$ versus with the parameter $M^2$ has been presented in Fig. 1, where the shaded band shows the uncertainties from all the mentioned input parameters. As shown by Fig. 1, the slope of $f_{\eta_s}$ is $\simeq 0.0085$, indicating the nearly flatness of the decay constant within the allowable Borel window. Using the above criteria, we obtain $M^2 \in [1.4, 2.0]$. Our results of $f_{\eta_s}$ at scale 1 GeV are presented in Table II, where as a comparison several typical results are also presented. It shows that our prediction is in good agreement with the result derived by using decay widths of $\eta \to \gamma\gamma$ and $\eta' \to \gamma\gamma$ under the QF mixing scheme [27], the previous QCDSR result [80] and the lattice QCD results [81–83].[1]

Similarly, we determine the suitable Borel windows for the moments $\langle\xi_{n;\eta_s}\rangle$ by requiring the continuum contribution and the contribution of the dimension-six condensates to be small. We list the first four moments $\langle\xi_{n;\eta_s}\rangle$ at the scale $\mu = \sqrt{M^2}$ in Table III, where contributions of the continuum and dimension-six condensates together with the determined Borel windows are also presented. Then the moments $\langle\xi_{n;\eta_s}\rangle$ and the Gegenbauer moments $a_{n;\eta_s}$ for $n = 2, 4, 6$ which solved by Eqs. (38)–(40) at the initial scale $\mu_0 = 1$ GeV are $\langle\xi_{2;\eta_s}\rangle = 0.194 \pm 0.008$, $\langle\xi_{4;\eta_s}\rangle = 0.087 \pm 0.006$, $\langle\xi_{6;\eta_s}\rangle = 0.051 \pm 0.007$, $a_{2;\eta_s} = -0.017 \pm 0.023$, $a_{4;\eta_s} = 0.052^{+0.003}_{-0.004}$, and $a_{6;\eta_s} = -0.011 \pm 0.097$, respectively.

[1]Note that there is $\sqrt{2}$ difference for our adopted definition of pseudoscalar decay constant with that of Ref. [82], so the results of Ref. [82] need to be multiplied by $\sqrt{2}$.

TABLE II. The decay constant $f_{\eta_s}$ using the BFTSR approach compared with other typical results.

| References | $f_{\eta_s}$ [GeV] |
|---|---|
| This work (BFTSR) | $0.176^{+0.006}_{-0.005}$ |
| QCDSR 2000 [80] | $0.178^{+0.004}_{-0.004}$ |
| Feldmann and Kroll 1998 [27] | $0.176^{+0.008}_{-0.008}$ |
| The Effective Lagrangian 2015 [84] | $0.144^{+0.006}_{-0.006}$ |
| LQCD 2018 [81] | $0.178^{+0.004}_{-0.004}$ |
| LQCD 2021 [82] | $0.179^{+0.006}_{-0.006}$ |
| LQCD 2025 [83] | $0.171^{+0.003}_{-0.003}$ |

TABLE III. The determined Borel windows and the $\eta_s$ leading-twist LCDA moments $\langle\xi_{n;\eta_s}\rangle$ at the scale $\mu = \sqrt{M^2}$. "Con." represents the continuum contribution and "Six." is the contribution of dimension-six condensates.

| | $n = 2$ | $n = 4$ | $n = 6$ |
|---|---|---|---|
| Con. | $< 35\%$ | $< 40\%$ | $< 45\%$ |
| Six. | $< 5\%$ | $< 5\%$ | $< 5\%$ |
| $M^2$ | [1.026, 2.470] | [1.369, 2.925] | [1.677, 3.520] |
| $\langle\xi_{n;\eta_s}\rangle\vert_\mu$ | [0.193, 0.198] | [0.083, 0.090] | [0.046, 0.054] |

### C. Mixing parameters and the related masses

As for the mixing parameters of the most complex four-particle mixing, we fix their values via a step-by-step way, e.g., by using the values determined from the $\eta-\eta'$, $\eta-\eta'-\eta_c$, and $\eta-\eta'-G$ mixings as the starting point, recursively, so as to fix all the required values for the most complex case of $\eta-\eta'-G-\eta_c$. Considering the OZI-rule violating and suppression, the decay constants $f^q_{s,g,c}$, $f^s_{q,g,c}$ and $f^c_{q,s,g}$ and the off-diagonal pseudoscalar density matrix elements $m^2_{qs}$, $m^2_{qg}$, $m^2_{qc}$, and etc. are small and can be safely set as zeros. Using these approximations, the mixing of $\eta-\eta'-G-\eta_c$ involves twelve parameters, e.g., four mass-related terms $m^2_{qq}$, $m^2_{ss}$, $m^2_{cc}$, and $m_G$, four U(1)-anomaly matrix elements $G_q$, $G_s$, $G_c$, and $G_g$, and four mixing angles $\theta$, $\phi_g$, $\phi_c$, and $\theta_g$, respectively.

First, by using the case of $\eta-\eta'$ mixing under the QF scheme, we can fix four parameters $m^2_{qq}$, $m^2_{ss}$, $G_q$, and $G_s$ by using their following relations to the given parameters $m^2_\eta$, $m^2_{\eta'}$, $f_q$, $f_s$, and $\phi$:

$$m^2_\eta \cos^2\phi + m^2_{\eta'} \sin^2\phi = m^2_{qq} + \frac{\sqrt{2}}{f_q} G_q, \tag{54}$$

$$(m_{\eta'}^2 - m_\eta^2)\sin\phi\cos\phi = \frac{1}{f_s}G_q, \tag{55}$$

$$(m_{\eta'}^2 - m_\eta^2)\sin\phi\cos\phi = \frac{\sqrt{2}}{f_q}G_s, \tag{56}$$

$$m_\eta^2\sin^2\phi + m_{\eta'}^2\cos^2\phi = m_{ss}^2 + \frac{1}{f_s}G_s. \tag{57}$$

The mixing angle $\phi$ has been studied by various groups, whose magnitude varies for different methods. For clarity, we take the recent value $\phi = (41.2^{+0.05}_{-0.06})^\circ$ determined by using the QCD sum rules approach [66] to do our discussion. After using the values of parameter sets in Sec. III A, we obtain

$$m_{qq}^2 = 0.029^{+0.025}_{-0.026}\ \mathrm{GeV}^2, \tag{58}$$

$$m_{ss}^2 = 0.476^{+0.008}_{-0.008}\ \mathrm{GeV}^2, \tag{59}$$

$$G_q = 0.054^{+0.002}_{-0.002}\ \mathrm{GeV}^3, \tag{60}$$

$$G_s = 0.030^{+0.001}_{-0.001}\ \mathrm{GeV}^3. \tag{61}$$

Second, we can fix the values of $m_{cc}^2$ and $G_c$ by using the case of $\eta-\eta'-\eta_c$ mixing with the inputs $m_{nc} = 2.9841 \pm 0.0004$ GeV [13] and $f_c = 0.453 \pm 0.004$ GeV [85], and the values of $G_g$ and the pseudoscalar glueball mass $m_G$ by using the case of $\eta-\eta'-G$ mixing. Our results are

$$m_{cc}^2 = 8.876^{+0.003}_{-0.003}\ \mathrm{GeV}^2, \tag{62}$$

$$G_c = 0.012^{+0.001}_{-0.001}\ \mathrm{GeV}^3, \tag{63}$$

$$m_G = 1.509^{+0.133}_{-0.130}\ \mathrm{GeV}, \tag{64}$$

$$G_g = 0.000^{+0.008}_{-0.007}\ \mathrm{GeV}^3, \tag{65}$$

Here we have adopted the mixing angles $\theta = (-13.5^{+3.070}_{-2.762})^\circ$ and $\phi_g = (-0.132^{+0.599}_{-0.398})^\circ$ that are fixed in the $\eta-\eta'-G$ mixing case to do the fixing. As shown by Sec. II A and with the help of Eqs. (A14) and (A15), the glueball mass $m_G$ can be related to quantities such as the $\eta$ and $\eta'$ masses, the decay constants, and the mixing angles via the following relation,

$$\frac{m_\eta^2 s\theta s\phi_G(s\theta s\theta_i c\phi_G - c\theta c\theta_i) + m_{\eta'}^2 c\theta s\phi_G(c\theta s\theta_i c\phi_G + s\theta c\theta_i) - m_G^2 s\theta_i s\phi_G c\phi_G}{m_\eta^2 s\theta s\phi_G(s\theta c\theta_i c\phi_G + c\theta s\theta_i) + m_{\eta'}^2 c\theta s\phi_G(c\theta c\theta_i c\phi_G - s\theta s\theta_i) - m_G^2 c\theta_i s\phi_G c\phi_G} = \frac{\sqrt{2}f_s}{f_q}. \tag{66}$$

It is noted that our value of the pseudoscalar glueball mass $m_G$ is lower than the most recent BES III measurements [86], which indicate that $X(2370)$ is likely a good candidate of pseudoscalar glueball, cf., the most recent review [87]. Our value is however consistent with the value $m_G = 1.4 \pm 0.1$ GeV given by Refs [35,63] and is consistent with the result $m_G = 1.75 \pm 0.16$ GeV [88] calculated by using inverse matrix method to do a dispersive analysis on the pseudoscalar glueball mass. It is noted that the predicted glueball mass will increase with the increment of $f_q$ and the decrement of $f_s$.[2] Thus, further studies are still needed to clarify this issue.

Furthermore, mapping the matrix elements from matrix (17) to Eq. (A16) and taking the mixing parameters fixed above as inputs, we derive the mixing angles in the $\eta-\eta'-G-\eta_c$ mixing case with $\theta = (-13.441)^\circ$, $\theta_g = (-0.005)^\circ$, $\phi_c = (-1.213)^\circ$ and $\phi_g = (-0.133)^\circ$. It is important to note that the mixing angle $\theta$ is slightly modified from which in the $\eta-\eta'-G$ mixing case due to the inclusion of $\eta_c$. At this point, all the parameters involved in the $\eta-\eta'-G-\eta_c$ case have been obtained. Using these parameters, we then obtain the final matrix for the mixing pattern of $\eta-\eta'-G-\eta_c$ as follows:

$$U(\theta,\phi_g,\phi_c,\theta_g) = \begin{pmatrix} 0.7517 & -0.6595 & -0.0005 & -0.0049 \\ 0.6593 & 0.7516 & -0.0023 & -0.0206 \\ 0.0019 & 0.0013 & 1.000 & -0.0001 \\ 0.0173 & 0.0122 & 0.0001 & 0.9998 \end{pmatrix}. \tag{67}$$

From the above mixing matrix, we find that the mixing of gluon and $\eta$, $\eta'$-meson is relatively small, which is consistent with the previous observation of Ref. [89].

### D. The $H \to \eta^{(\prime)}$ TFFs

In QF scheme, the main contribution of $H \to \eta^{(\prime)}$ TFFs comes from the $|\eta_q\rangle$-component for $H = B^+, D^+$ mesons and from the $|\eta_s\rangle$-component for $H = D_s^+$ meson since the $s\bar{s}$ component can be accessed only via $D_s$ meson decay

[2]Following the BFTSR method, we have shown that $f_q = 0.141 \pm 0.005$ GeV and $f_s = 0.176^{+0.006}_{-0.005}$ GeV. It is found that by fixing $f_q$ within its determined range but taking a smaller $f_s$ outside the presently determined range, we can obtain a larger pseudoscalar glueball mass closer to that of $X(2370)$. For example, if we take $f_q = 0.141$ GeV and $f_s \simeq 0.151$ GeV, or $f_q = 0.146$ GeV and $f_s \simeq 0.156$ GeV, we obtain $m_G \simeq 2.3$ GeV.

[48]. And we take the above mixing results (67) to derive the required $H\to\eta^{(\prime)}$ TFFs.

Following the usual choice, we set the continuum threshold $s_0^{H\to\eta^{(\prime)}}$ to be near the squared mass of the first excited state of the $B$, $D$, and $D_s$ mesons. The continuum threshold $s_0^{H\to\eta^{(\prime)}}$ and the Borel parameters for different decay processes are

$$
\begin{aligned}
s_0^{B\to\eta} &= 37.0\pm1.0\ \mathrm{GeV}^2, & M^2_{B\to\eta} &= 20.0\pm2.0\ \mathrm{GeV}^2,\\
s_0^{B\to\eta'} &= 36.0\pm1.0\ \mathrm{GeV}^2, & M^2_{B\to\eta'} &= 20.0\pm2.0\ \mathrm{GeV}^2,\\
s_0^{D\to\eta} &= 7.0\pm0.3\ \mathrm{GeV}^2, & M^2_{D\to\eta} &= 8.0\pm1.0\ \mathrm{GeV}^2,\\
s_0^{D\to\eta'} &= 7.0\pm0.3\ \mathrm{GeV}^2, & M^2_{D\to\eta'} &= 8.0\pm1.0\ \mathrm{GeV}^2,\\
s_0^{D_s\to\eta} &= 7.5\pm0.2\ \mathrm{GeV}^2, & M^2_{D_s\to\eta} &= 20.0\pm1.0\ \mathrm{GeV}^2,\\
s_0^{D_s\to\eta'} &= 7.8\pm0.2\ \mathrm{GeV}^2, & M^2_{D_s\to\eta'} &= 25.0\pm2.0\ \mathrm{GeV}^2.
\end{aligned}
\tag{68}
$$

Here the Borel windows are fixed by using usual criteria, e.g., (1) The contribution of the continuum states to the total TFFs is less than 30%; (2) The contribution of the twist-4 LCDA to the total TFFs is less than 5%; and (3) The TFFs are stable within the Borel window.

We present our predictions of the TFFs at $q^2=0$ in Table IV, whose errors are squared averages of the ones from the mentioned input parameters. As a comparison, the experimental results of BES-III Collaboration [90–92] and typical results derived by using various approaches, such as the LCSR approach [38,39,49,66,67], the pQCD approach [93], the covariant light front (CLF) approach [94], the covariant confining quark mode (CCQM) approach [95], are also presented. Our results of $f^+_{B\to\eta'}(0)$, $f^+_{D\to\eta'}(0)$, and $f^+_{D_s\to\eta'}(0)$ are closer to those of $f^+_{B\to\eta}(0)$, $f^+_{D\to\eta}(0)$, and $f^+_{D_s\to\eta}(0)$ than previous LCSR predictions. The main reason for this is due to the different twist-3 parameters $h_q$ and $h_s$ with previous choices are adopted for our calculation, e.g., Refs. [38,39,49,66,67] directly adopted the values $h_q=0.0015\pm0.0040$ and $h_s=0.087\pm0.006\ \mathrm{GeV}^2$ of Ref. [96][3] to do the calculation. At present, we obtain $h_q=0.00405^{+0.00367}_{-0.00373}\ \mathrm{GeV}^3$ and $h_s=0.08368^{+0.00379}_{-0.00366}\ \mathrm{GeV}^3$ by using Eqs. (52) and (53). As shown by Fig. 2, the twist-3 contributions are positive and sizable, and their contributions to the TFFs $f^+_{B\to\eta'}(0)$, $f^+_{D\to\eta'}(0)$, and $f^+_{D_s\to\eta'}(0)$ are then more sensitive to the magnitudes of $h_q$ and $h_s$, especially for the cases of $B/D$ mesons.

[3] In this reference, the leading-order approximation of $f_{\eta_q}=f_\pi$ and $f_{\eta_s}=1.41f_\pi$ is adopted for calculating Eqs. (52) and (53). While our magnitudes of those two decay constants are calculated by using the LCSR approach.

TABLE IV. Typical theoretical predictions and experimental data on the TFFs $f^+_{H\to\eta^{(\prime)}}(0)$ at the large recoil point $q^2=0$.

| | $f^+_{B\to\eta}(0)$ | $f^+_{B\to\eta'}(0)$ |
|---|---|---|
| This work | $0.147^{+0.013}_{-0.011}$ | $0.140^{+0.025}_{-0.023}$ |
| LCSR 2023 [67] | $0.145^{+0.009}_{-0.010}$ | $0.128^{+0.008}_{-0.009}$ |
| LCSR 2015 [39] | $0.168^{+0.041}_{-0.047}$ | $0.130^{+0.036}_{-0.032}$ |
| LCSR 2013 [49] | $0.238^{+0.046}_{-0.046}$ | $0.198^{+0.039}_{-0.039}$ |
| LCSR 2007 [38] | $0.229^{+0.035}_{-0.035}$ | $0.188^{+0.028}_{-0.028}$ |
| pQCD 2006 [93] | 0.147 | 0.121 |
| CLF 2009 [94] | $0.220^{+0.018}_{-0.018}$ | $0.180^{+0.016}_{-0.016}$ |
| | $f^+_{D\to\eta}(0)$ | $f^+_{D\to\eta'}(0)$ |
| This work | $0.336^{+0.038}_{-0.039}$ | $0.339^{+0.061}_{-0.061}$ |
| LCSR 2023 [67] | $0.329^{+0.021}_{-0.015}$ | $0.294^{+0.021}_{-0.015}$ |
| LCSR 2015 [39] | $0.429^{+0.165}_{-0.141}$ | $0.292^{+0.113}_{-0.104}$ |
| LCSR 2013 [49] | $0.552^{+0.051}_{-0.051}$ | $0.458^{+0.105}_{-0.105}$ |
| BES-III 2025 [90] | $0.345^{+0.008+0.003}_{-0.008-0.003}$ | … |
| BES-III 2020 [91] | $0.39^{+0.04+0.01}_{-0.04-0.01}$ | … |
| CCQM 2019 [95] | $0.36^{+0.05}_{-0.05}$ | $0.36^{+0.05}_{-0.05}$ |
| | $f^+_{D_s\to\eta}(0)$ | $f^+_{D_s\to\eta'}(0)$ |
| This work | $0.522^{+0.059}_{-0.057}$ | $0.548^{+0.081}_{-0.079}$ |
| LCSR 2021 [66] | $0.476^{+0.040}_{-0.036}$ | $0.544^{+0.046}_{-0.042}$ |
| LCSR 2015 [39] | $0.495^{+0.030}_{-0.029}$ | $0.557^{+0.048}_{-0.045}$ |
| LCSR 2013 [49] | $0.432^{+0.033}_{-0.033}$ | $0.520^{+0.080}_{-0.080}$ |
| BES-III 2024 [92] | $0.482^{+0.011+0.009+0.004}_{-0.011-0.009-0.004}$ | $0.562^{+0.031+0.014+0.003}_{-0.031-0.014-0.003}$ |
| CCQM 2019 [95] | $0.49^{+0.07}_{-0.07}$ | $0.59^{+0.09}_{-0.09}$ |
| LFQM 2009 [97] | 0.50 | 0.62 |

Figure 2 shows how the TFFs $f^+_{H\to\eta^{(\prime)}}(q^2)$ vary with the increment of $q^2$, where the total and separate contributions from different twist LCDAs are given. Figure 2 shows that the twist-2 terms give dominant contributions to the TFFs, the twist-3 contributions are sizable, and the twist-4 contributions are negligible. The TFFs $f^+_{H\to\eta^{(\prime)}}(0)$ with the uncertainties caused by different inputs at the large recoil point are arranged as follows:

$$
\begin{aligned}
f^+_{B\to\eta}(0) &= 0.147\left(^{+0.004}_{-0.004}\right)_{s_0}\left(^{+0.001}_{-0.001}\right)_{M^2}\left(^{+0.009}_{-0.006}\right)_{a^\eta_{2;g}}\left(^{+0.008}_{-0.008}\right)_{\mathrm{rest}}\\
&= 0.147^{+0.013}_{-0.011},
\end{aligned}
\tag{69}
$$

$$
\begin{aligned}
f^+_{B\to\eta'}(0) &= 0.140\left(^{+0.004}_{-0.004}\right)_{s_0}\left(^{+0.002}_{-0.001}\right)_{M^2}\left(^{+0.023}_{-0.021}\right)_{a^{\eta'}_{2;g}}\left(^{+0.007}_{-0.007}\right)_{\mathrm{rest}}\\
&= 0.140^{+0.025}_{-0.023},
\end{aligned}
\tag{70}
$$

$$
\begin{aligned}
f^+_{D\to\eta}(0) &= 0.336\left(^{+0.005}_{-0.006}\right)_{s_0}\left(^{+0.0004}_{-0.0004}\right)_{M^2}\left(^{+0.032}_{-0.033}\right)_{a^\eta_{2;g}}\left(^{+0.020}_{-0.019}\right)_{\mathrm{rest}}\\
&= 0.336^{+0.038}_{-0.039},
\end{aligned}
\tag{71}
$$

$$
\begin{aligned}
f^+_{D\to\eta'}(0) &= 0.339\left(^{+0.006}_{-0.006}\right)_{s_0}\left(^{+0.0005}_{-0.0004}\right)_{M^2}\left(^{+0.057}_{-0.058}\right)_{a^{\eta'}_{2;g}}\left(^{+0.020}_{-0.019}\right)_{\mathrm{rest}}\\
&= 0.339^{+0.061}_{-0.061},
\end{aligned}
\tag{72}
$$

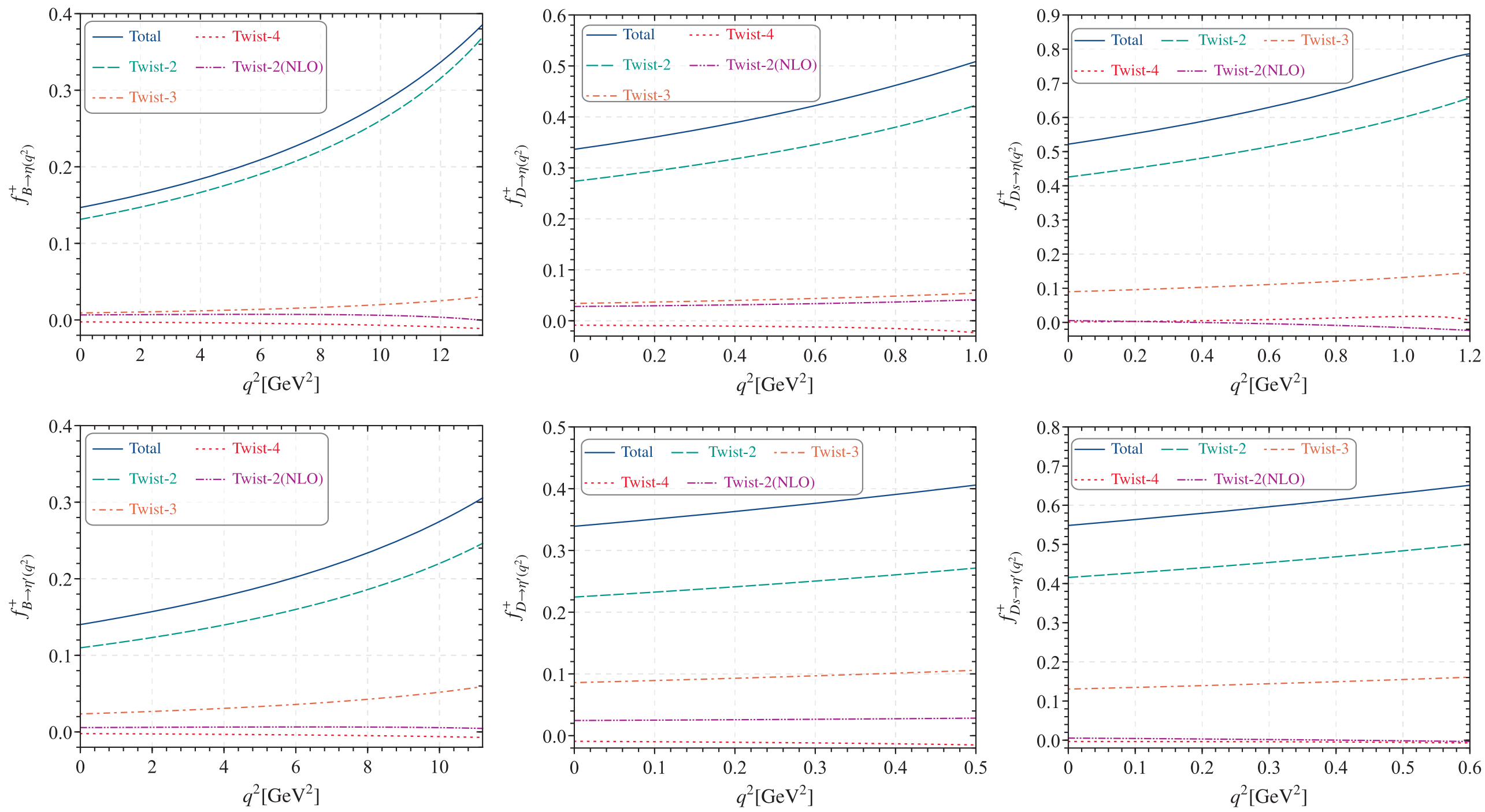

FIG. 2. The LCSR predictions on the TFFs $f^+_{H\to\eta^{(\prime)}}(q^2)$ with $H = B^+$, $D^+$ and $D_s^+$, where the contributions from the twist-2, twist-3, and twist-4 LCDAs are given separately. The twist-2 terms are given up to NLO QCD corrections.

$$f^+_{D_s\to\eta}(0) = 0.522(^{+0.006}_{-0.006})_{s_0}(^{+0.00005}_{-0.00004})_{M^2}(^{+0.039}_{-0.041})_{a^{\eta}_{2;g}}(^{+0.043}_{-0.038})_{\rm rest} = 0.522^{+0.059}_{-0.057}, \quad (73)$$

$$f^+_{D_s\to\eta'}(0) = 0.548(^{+0.007}_{-0.007})_{s_0}(^{+0.0004}_{-0.0005})_{M^2}(^{+0.066}_{-0.068})_{a^{\eta'}_{2;g}}(^{+0.045}_{-0.040})_{\rm rest} = 0.548^{+0.081}_{-0.079}, \quad (74)$$

where rest represents the error caused by the other input parameters such as those of $m_b$, $f_{\eta_{q,s}}$, $f_{B,D,D_s}$ and the Gegenbauer moments of $\eta_{q,s}$. The errors are mainly caused by the Gegenbauer moments $a^{\eta^{(\prime)}}_{2;g}$, which varies greatly. The QCD sum rules 1999 determined $a^{\eta^{(\prime)}}_{2;g}(1\ \mathrm{GeV}) = 0.2$ [98]. The perturbative QCD 2013 fitted the data from the *BABAR* and the CLEO and got $a^{\eta^{(\prime)}}_{2;g}(1\ \mathrm{GeV}) = 19 \pm 5$ [75]. A fit of the CLEO and L3 data on the $\eta - \gamma$ and $\eta' - \gamma$ transition form factors provides $a^{\eta^{(\prime)}}_{2;g}(1\ \mathrm{GeV}) = 9 \pm 12$ [36]. And combining the coefficients from an analysis of the $\eta - \gamma$ and $\eta' - \gamma$ TFFs with the requirement of the positivity of the effective vertex function leads to $a^{\eta^{(\prime)}}_{2;g}(2\ \mathrm{GeV}) = 4.6 \pm 2.5$ [99]. So considering the large uncertainties about their values, in the calculation we take a very conservative range $a^{\eta}_{2;g} = a^{\eta'}_{2;g} = 0 \pm 20$ which is the same as the treatment of Refs. [38,39].

Additionally, we also calculate the TFFs at the large recoil point by using the mentioned values of the Gegenbauer moments $a^{\eta^{(\prime)}}_{2;g}$ and put the results in Table V. It shows that the total TFFs exhibit a positive correlation with the Gegenbauer moment, e.g. their values increases with the increment of the Gegenbauer moments. If we be able to determine a more precise value for the Gegenbauer moment, it would enable us to accurately calculate the gluon contribution.

Figure 3 depicts the gluonic contribution to the TFFs $f^+_{H\to\eta^{(\prime)}}(q^2)$ with $a^{\eta^{(\prime)}}_{2;g} = 0 \pm 20$. From these pictures we can see that the uncertainties of the gluonic contributions to $f^+_{H\to\eta^{(\prime)}}(q^2)$ from the errors of $a^{\eta^{(\prime)}}_{2;g}$ are much larger in $B/D/D_s \to \eta'$ decays than $B/D/D_s \to \eta$ decays. This is consistent with the conclusion drawn in Refs. [7,38,93]. The gluonic contribution to the TFFs are sizable. For the case of $a^{\eta^{(\prime)}}_{2;g} = 20$, at the large recoil point, the gluonic contributions are 6.0%, 16.6%, 9.5%, 16.9%, 7.5%, 12.1% for $f^+_{B\to\eta}(0)$, $f^+_{B\to\eta'}(0)$, $f^+_{D\to\eta}(0)$, $f^+_{D\to\eta'}(0)$, $f^+_{D_s\to\eta}(0)$, and $f^+_{D_s\to\eta'}(0)$, respectively.

The physically allowable ranges of the above heavy-to-light TFFs are $m_\ell^2 \leq q^2 \leq (m_{B^+/D^+/D_s^+} - m_{\eta^{(\prime)}})^2$. The LCSR approach is applicable only in the low and intermediate region, which needs to be extrapolated into whole $q^2$ region via proper extrapolation approaches. There are many fitting methods to extend the TFFs to whole $q^2$ region, such as the simple pole model [100], the Becirevic-Kaidalov (BK) parametrization [101], the double-pole parametrization [102], the Boyd-Grinstein-Lebed (BGL) parametrization [103], the converging simplified series expansion (SSE) [104,105], and etc. Given that the SSE parametrization offers a notable advantage by effectively converting the near-threshold behavior of the TFFs into a restrictive condition on the expansion coefficients, we will utilize the SSE method to do the extrapolation, thereby proposing a straightforward parametrization for the TFFs,

$$f^+_{H\to\eta^{(\prime)}}(q^2)=\frac{1}{1-q^2/m^2_{R^*}}\sum_k b_k z^k(t,t_0), \qquad (75)$$

where $m_{B^*}=5.3248$ GeV, $m_{D^*}=2.0103$ GeV, and $m_{D_s^*}=2.1122$ GeV [13] are the masses of resonance vector mesons and $z(t,t_0)$ is a function with the following form

$$z(t,t_0)=\frac{\sqrt{t_+-t}-\sqrt{t_+-t_0}}{\sqrt{t_+-t}+\sqrt{t_+-t_0}}, \qquad (76)$$

with $t_\pm=(m_H\pm m_{\eta^{(\prime)}})^2$, and $t_0=t_+(1-\sqrt{1-t_-/t_+})$ is a free parameter which can be optimized to reduce the maximum value of $|z(t,t_0)|$ in the physical range. The parameter $b_k$ can be determined by requiring $\Delta<1\%$, where the parameter $\Delta$ measures the quality of extrapolation and is defined as

$$\Delta=\frac{\sum_t|F_i(t)-F_i^{\rm fit}(t)|}{\sum_t|F_i(t)|}\times 100. \qquad (77)$$

TABLE V. The results of TFFs $f^+_{H\to\eta^{(\prime)}}(0)$ at the large recoil point $q^2=0$ with different $a^{\eta^{(\prime)}}_{2;g}$. Here the errors are combined errors from all the mentioned error sources.

| | $f^+_{B\to\eta}(0)$ | $f^+_{B\to\eta'}(0)$ |
|---|---|---|
| $a^{\eta^{(\prime)}}_{2;g}=0^{+20}_{-20}$ | $0.1468^{+0.0126}_{-0.0105}$ | $0.1402^{+0.0248}_{-0.0226}$ |
| $a^{\eta^{(\prime)}}_{2;g}=0.2$ | $0.1469^{+0.0091}_{-0.0087}$ | $0.1405^{+0.0086}_{-0.0083}$ |
| $a^{\eta^{(\prime)}}_{2;g}=4.6^{+2.5}_{-2.5}$ | $0.1487^{+0.0097}_{-0.0092}$ | $0.1470^{+0.0091}_{-0.0095}$ |
| $a^{\eta^{(\prime)}}_{2;g}=9^{+12}_{-12}$ | $0.1506^{+0.0114}_{-0.0104}$ | $0.1534^{+0.0146}_{-0.0183}$ |
| $a^{\eta^{(\prime)}}_{2;g}=19^{+5}_{-5}$ | $0.1548^{+0.0117}_{-0.0116}$ | $0.1681^{+0.0119}_{-0.0170}$ |
| | $f^+_{D\to\eta}(0)$ | $f^+_{D\to\eta'}(0)$ |
| $a^{\eta^{(\prime)}}_{2;g}=0^{+20}_{-20}$ | $0.3364^{+0.0379}_{-0.0389}$ | $0.3391^{+0.0610}_{-0.0611}$ |
| $a^{\eta^{(\prime)}}_{2;g}=0.2$ | $0.3368^{+0.0206}_{-0.0204}$ | $0.3398^{+0.0204}_{-0.0200}$ |
| $a^{\eta^{(\prime)}}_{2;g}=4.6^{+2.5}_{-4.5}$ | $0.3442^{+0.0235}_{-0.0239}$ | $0.3547^{+0.0230}_{-0.0249}$ |
| $a^{\eta^{(\prime)}}_{2;g}=9^{+12}_{-12}$ | $0.3517^{+0.0328}_{-0.0344}$ | $0.3697^{+0.0394}_{-0.0477}$ |
| $a^{\eta^{(\prime)}}_{2;g}=19^{+5}_{-5}$ | $0.3686^{+0.0389}_{-0.0400}$ | $0.4036^{+0.0359}_{-0.0449}$ |
| | $f^+_{D_s\to\eta}(0)$ | $f^+_{D_s\to\eta'}(0)$ |
| $a^{\eta^{(\prime)}}_{2;g}=0^{+20}_{-20}$ | $0.5217^{+0.0587}_{-0.0566}$ | $0.5484^{+0.0806}_{-0.0793}$ |
| $a^{\eta^{(\prime)}}_{2;g}=0.2$ | $0.5222^{+0.0437}_{-0.0388}$ | $0.5492^{+0.0460}_{-0.0407}$ |
| $a^{\eta^{(\prime)}}_{2;g}=4.6^{+2.5}_{-4.5}$ | $0.5313^{+0.0459}_{-0.0415}$ | $0.5664^{+0.0486}_{-0.0452}$ |
| $a^{\eta^{(\prime)}}_{2;g}=9^{+12}_{-12}$ | $0.5405^{+0.0539}_{-0.0515}$ | $0.5836^{+0.0623}_{-0.0665}$ |
| $a^{\eta^{(\prime)}}_{2;g}=19^{+5}_{-5}$ | $0.5613^{+0.0603}_{-0.0575}$ | $0.6227^{+0.0637}_{-0.0668}$ |

The fitting parameters $b_{1,2}$ for each TFFs and the quality-of-fit $\Delta$ in Table VI. It shows that under those choices of parameters, all the $\Delta$ values are no more than 0.880%, indicating a good agreement of the extrapolated curves with the LCSRs within the same $q^2$-region.

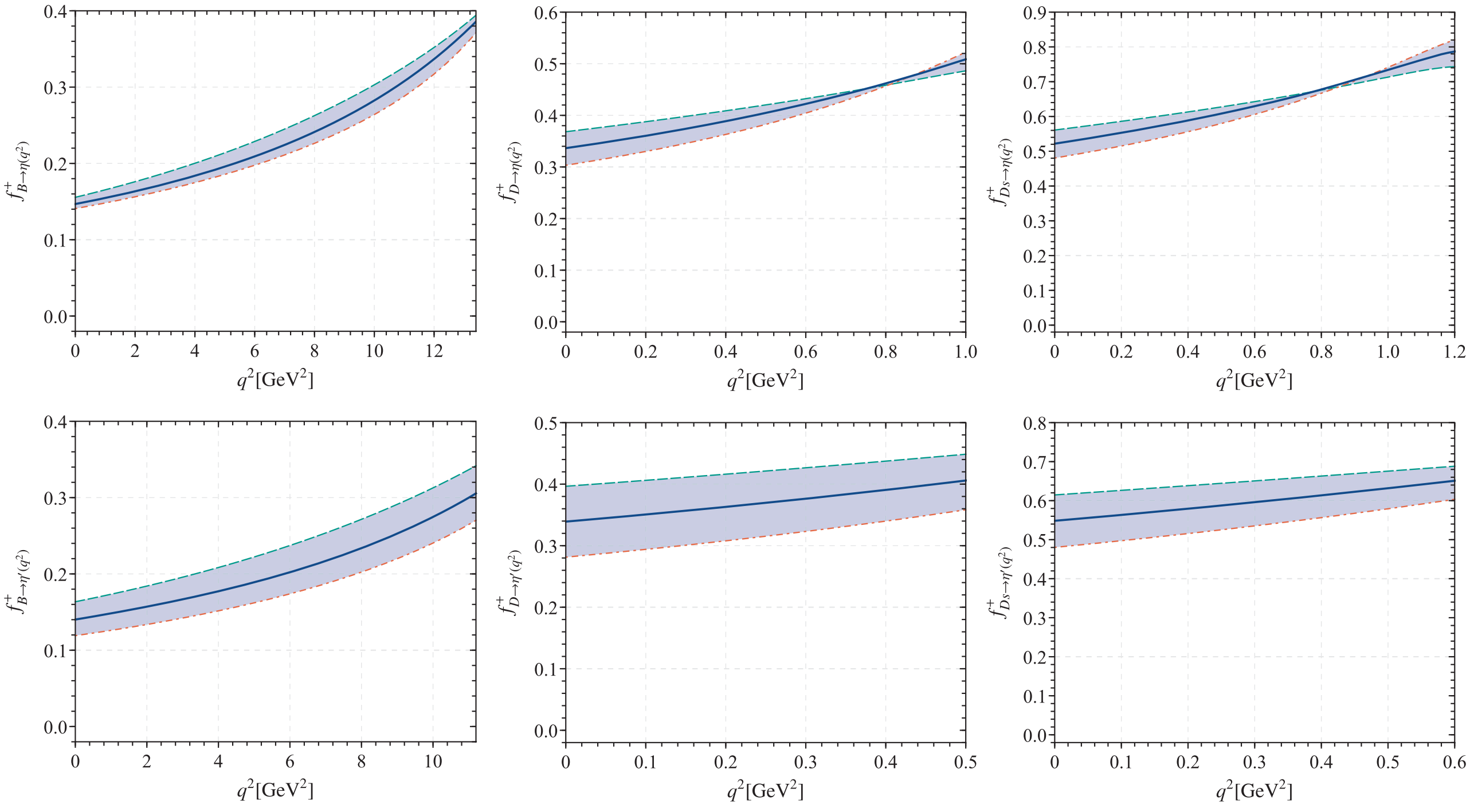

FIG. 3. The gluonic dependence of the TFFs $f^+_{H\to\eta^{(\prime)}}(q^2)$ with $H=B^+, D^+$, and $D_s^+$, respectively. The solid lines represent the result of $f^+_{H\to\eta^{(\prime)}}(q^2)$ with $a^{\eta^{(\prime)}}_{2;g}=0$, dashed lines with $a^{\eta^{(\prime)}}_{2;g}=20$ and dashed-dotted lines with $a^{\eta^{(\prime)}}_{2;g}=-20$. The shaded areas show the change of the TFFs under the variation of $a^{\eta^{(\prime)}}_{2;g}=0\pm 20$.

TABLE VI. Fitting results of $b_1$ and $b_2$ for the central TFFs $f^{+(c)}_{H\to\eta^{(\prime)}}(q^2)$, the upper TFFs $f^{+(u)}_{H\to\eta^{(\prime)}}(q^2)$, and the lower TFFs $f^{+(l)}_{H\to\eta^{(\prime)}}(q^2)$, respectively. $\Delta$ is the measure of SSE extrapolation quality.

| | $f^{+(c)}_{B\to\eta}(0)$ | $f^{+(u)}_{B\to\eta}(0)$ | $f^{+(l)}_{B\to\eta}(0)$ | $f^{+(c)}_{B\to\eta'}(0)$ | $f^{+(u)}_{B\to\eta'}(0)$ | $f^{+(l)}_{B\to\eta'}(0)$ |
|---|---|---|---|---|---|---|
| $b_1$ | −0.343 | −0.568 | −0.238 | −0.416 | −0.614 | −0.350 |
| $b_2$ | 0.858 | −1.007 | 1.338 | 1.231 | −1.030 | 1.763 |
| $\Delta$ | 0.177% | 0.767% | 0.413% | 0.103% | 0.880% | 0.177% |
| | $f^{+(c)}_{D\to\eta}(0)$ | $f^{+(u)}_{D\to\eta}(0)$ | $f^{+(l)}_{D\to\eta}(0)$ | $f^{+(c)}_{D\to\eta'}(0)$ | $f^{+(u)}_{D\to\eta'}(0)$ | $f^{+(l)}_{D\to\eta'}(0)$ |
| $b_1$ | −0.653 | −0.058 | −1.011 | −0.891 | 0.081 | −1.593 |
| $b_2$ | 7.019 | 5.859 | 13.205 | 8.455 | −2.486 | 22.491 |
| $\Delta$ | 0.067% | 0.552% | 0.576% | 0.001% | 0.002% | 0.003% |
| | $f^{+(c)}_{D_s\to\eta}(0)$ | $f^{+(u)}_{D_s\to\eta}(0)$ | $f^{+(l)}_{D_s\to\eta}(0)$ | $f^{+(c)}_{D_s\to\eta'}(0)$ | $f^{+(u)}_{D_s\to\eta'}(0)$ | $f^{+(l)}_{D_s\to\eta'}(0)$ |
| $b_1$ | −0.710 | 0.040 | −1.197 | −1.840 | −0.327 | −3.204 |
| $b_2$ | 9.351 | 22.077 | 6.331 | −75.293 | −61.896 | −86.963 |
| $\Delta$ | 0.134% | 0.173% | 0.419% | 0.442% | 0.323% | 0.679% |

### E. Branching fractions for the semileptonic decay $B^+/D^+/D_s^+\to\eta^{(\prime)}\ell^+\nu_\ell$

Figure 4 presents the different decay widths for $H\to\eta^{(\prime)}\ell^+\nu_\ell$ without CKM matrix elements. The total decay widths for $H\to\eta^{(\prime)}\ell^+\nu_\ell$ with two different channel $\Gamma(H\to\eta^{(\prime)}e^+\nu_e)$ and $\Gamma(H\to\eta^{(\prime)}\mu^+\nu_\mu)$ can be obtained by integrating over the whole $q^2$ region, $m_\ell^2\le q^2\le(m_H-m_{\eta^{(\prime)}})^2$. Our predicted values for these decay widths are as follows

$$\Gamma(B^+\to\eta e^+\nu_e)=1.761^{+0.329}_{-0.279}\times10^{-17}\ \mathrm{GeV}, \tag{78}$$

$$\Gamma(B^+\to\eta\mu^+\nu_\mu)=1.760^{+0.329}_{-0.279}\times10^{-17}\ \mathrm{GeV}, \tag{79}$$

$$\Gamma(B^+\to\eta' e^+\nu_e)=1.136^{+0.385}_{-0.305}\times10^{-17}\ \mathrm{GeV}, \tag{80}$$

$$\Gamma(B^+\to\eta'\mu^+\nu_\mu)=1.135^{+0.384}_{-0.305}\times10^{-17}\ \mathrm{GeV}, \tag{81}$$

$$\Gamma(D^+\to\eta e^+\nu_e)=5.834^{+0.950}_{-0.794}\times10^{-16}\ \mathrm{GeV}, \tag{82}$$

$$\Gamma(D^+\to\eta\mu^+\nu_\mu)=5.764^{+0.934}_{-0.778}\times10^{-16}\ \mathrm{GeV}, \tag{83}$$

$$\Gamma(D^+\to\eta' e^+\nu_e)=1.268^{+0.409}_{-0.363}\times10^{-16}\ \mathrm{GeV}, \tag{84}$$

$$\Gamma(D^+\to\eta'\mu^+\nu_\mu)=1.231^{+0.395}_{-0.351}\times10^{-16}\ \mathrm{GeV}, \tag{85}$$

$$\Gamma(D_s^+\to\eta e^+\nu_e)=36.744^{+7.383}_{-5.966}\times10^{-15}\ \mathrm{GeV}, \tag{86}$$

$$\Gamma(D_s^+\to\eta\mu^+\nu_\mu)=36.35^{+7.290}_{-5.885}\times10^{-15}\ \mathrm{GeV}, \tag{87}$$

$$\Gamma(D_s^+\to\eta' e^+\nu_e)=10.13^{+2.652}_{-2.336}\times10^{-15}\ \mathrm{GeV}, \tag{88}$$

$$\Gamma(D_s^+\to\eta'\mu^+\nu_\mu)=9.884^{+2.575}_{-2.270}\times10^{-15}\ \mathrm{GeV}. \tag{89}$$

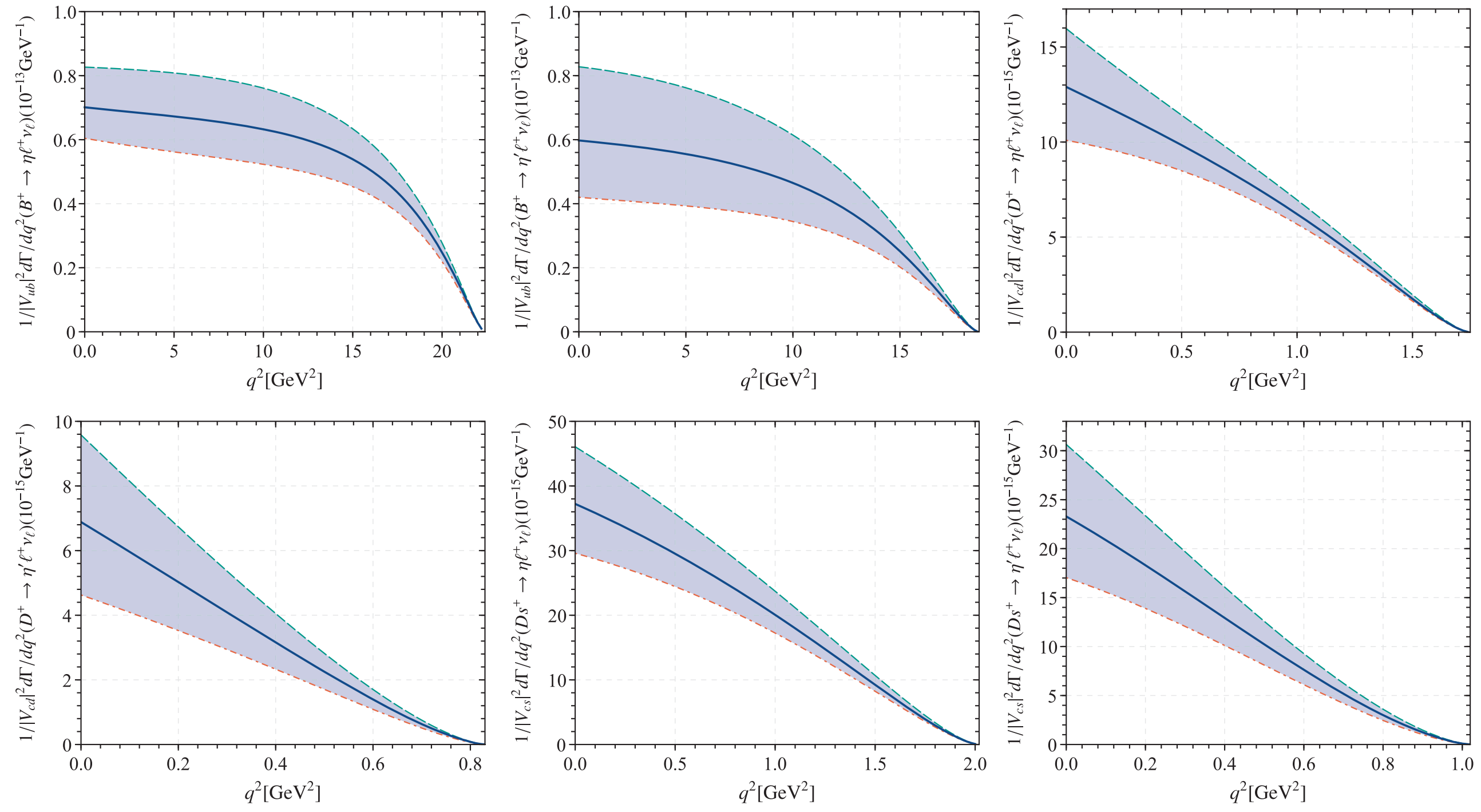

FIG. 4. The decay widths for $H\to\eta^{(\prime)}\ell^+\nu_\ell$, where the solid lines represent the central values and the shaded areas show their uncertainties, whose magnitudes are squared averages of all the mentioned error sources. $H$ represents $B^+$, $D^+$, and $D_s^+$, respectively.

TABLE VII. Branching fractions for the different decays integrated over the total $q^2$-region.

| | $\mathcal{B}(B^+ \to \eta e^+\nu_e)$ | $\mathcal{B}(B^+ \to \eta\mu^+\nu_\mu)$ | $\mathcal{B}(B^+ \to \eta' e^+\nu_e)$ | $\mathcal{B}(B^+ \to \eta'\mu^+\nu_\mu)$ |
|---|---|---|---|---|
| This work | $(4.382^{+0.818}_{-0.695}) \times 10^{-5}$ | $(4.379^{+0.818}_{-0.695}) \times 10^{-5}$ | $(2.827^{+0.957}_{-0.759}) \times 10^{-5}$ | $(2.825^{+0.956}_{-0.758}) \times 10^{-5}$ |
| PDG [13] | $(3.5^{+0.4}_{-0.4}) \times 10^{-5}$ | $(3.5^{+0.4}_{-0.4}) \times 10^{-5}$ | $(2.4^{+0.7}_{-0.7}) \times 10^{-5}$ | $(2.4^{+0.7}_{-0.7}) \times 10^{-5}$ |
| LCSR 2001 [106] | $(4.32^{+0.83}_{-0.83}) \times 10^{-5}$ | $(4.32^{+0.83}_{-0.83}) \times 10^{-5}$ | $(2.10^{+0.40}_{-0.40}) \times 10^{-5}$ | $(2.10^{+0.40}_{-0.40}) \times 10^{-5}$ |
| Belle 2022 [107] | $(2.83^{+0.55+0.34}_{-0.55-0.34}) \times 10^{-5}$ | $(2.83^{+0.55+0.34}_{-0.55-0.34}) \times 10^{-5}$ | $(2.79^{+1.29+0.30}_{-1.29-0.30}) \times 10^{-5}$ | $(2.79^{+1.29+0.30}_{-1.29-0.30}) \times 10^{-5}$ |
| | $\mathcal{B}(D^+ \to \eta e^+\nu_e)$ | $\mathcal{B}(D^+ \to \eta\mu^+\nu_\mu)$ | $\mathcal{B}(D^+ \to \eta' e^+\nu_e)$ | $\mathcal{B}(D^+ \to \eta'\mu^+\nu_\mu)$ |
| This work | $(0.916^{+0.149}_{-0.125}) \times 10^{-3}$ | $(0.905^{+0.147}_{-0.122}) \times 10^{-3}$ | $(1.990^{+0.642}_{-0.570}) \times 10^{-4}$ | $(1.932^{+0.619}_{-0.551}) \times 10^{-4}$ |
| PDG [13] | $(1.11^{+0.07}_{-0.07}) \times 10^{-3}$ | $(1.04^{+0.11}_{-0.11}) \times 10^{-3}$ | $(2.0^{+0.4}_{-0.4}) \times 10^{-4}$ | … |
| BES-III 2025 [90] | $(0.975^{+0.29+0.28}_{-0.29-0.28}) \times 10^{-3}$ | $(0.908^{+0.35+0.23}_{-0.35-0.23}) \times 10^{-3}$ | … | … |
| BES-III 2024 [108] | … | … | $(1.79^{+0.19+0.07}_{-0.19-0.07}) \times 10^{-4}$ | $(1.92^{+0.28+0.08}_{-0.28-0.08}) \times 10^{-4}$ |
| BES-III 2020 [91] | … | $(1.04^{+0.1+0.05}_{-0.1-0.05}) \times 10^{-3}$ | … | … |
| | $\mathcal{B}(D_s^+ \to \eta e^+\nu_e)$ | $\mathcal{B}(D_s^+ \to \eta\mu^+\nu_\mu)$ | $\mathcal{B}(D_s^+ \to \eta' e^+\nu_e)$ | $\mathcal{B}(D_s^+ \to \eta'\mu^+\nu_\mu)$ |
| This work | $(2.798^{+0.555}_{-0.448}) \times 10^{-2}$ | $(2.768^{+0.562}_{-0.454}) \times 10^{-2}$ | $(7.714^{+2.019}_{-1.779}) \times 10^{-3}$ | $(0.753^{+0.196}_{-0.173}) \times 10^{-2}$ |
| PDG [13] | $(2.26^{+0.06}_{-0.06}) \times 10^{-2}$ | $(2.4^{+0.05}_{-0.05}) \times 10^{-2}$ | $(8.0^{+0.4}_{-0.4}) \times 10^{-3}$ | $(1.1^{+0.5}_{-0.5}) \times 10^{-2}$ |
| LCSR 2021 [66] | $(2.346^{+0.418}_{-0.331}) \times 10^{-2}$ | $(2.320^{+0.413}_{-0.327}) \times 10^{-2}$ | $(7.92^{+1.41}_{-1.18}) \times 10^{-3}$ | $(0.773^{+0.138}_{-0.115}) \times 10^{-2}$ |
| BES-III 2024 [109] | … | $(2.235^{+0.051+0.052}_{-0.051-0.052}) \times 10^{-2}$ | … | $(0.801^{+0.055+0.028}_{-0.055-0.028}) \times 10^{-2}$ |

According to Eq. (30), using the decay lifetimes and the CKM matrix elements given by the PDG and the decay widths predicted by Eq. (29), we are ready to derive the branching fractions. We put our results and some typical measured values in Table VII. Our results are consistent with the PDG values within errors. Table VII also shows that our predictions for $D^+(D_s^+) \to \eta^{(\prime)}\ell\nu_\ell$ fall within the recent BES III measurements within errors. Hopefully, the decays $B^+ \to \eta^{(\prime)}\ell\nu_\ell$ can be observed in near future such as the Belle II, which inversely could provide a potential test for QCD sum rules approach.

## IV. SUMMARY

In this paper, we have calculated the first three Gegenbauer moments $a^{2,4,6}_{2;\eta_s}$ of the leading-twist LCDA and the decay constant $f_{\eta_s}$ of the $\eta_s$-state in the QF scheme using the QCD sum rules within the background field. This, together with our previous results for the $\eta_q$-state also in the QF scheme, can be used to calculate the heavy-to-light TFFs and the required mixed parameters for the FKS scheme more precisely.

Then we have constructed the $\eta-\eta'-G-\eta_c$ mixing based on the FKS scheme, where $G$ denotes the possible pseudoscalar glueball content. Implementing this mixing scheme into the equations of motion for the anomalous Ward identity, that connects the vacuum to $\eta$, $\eta'$, $\eta_c$, and $G$ transition matrix elements of the divergence of axial-vector currents to those of pseudoscalar densities and the U(1) anomaly, we have obtained the relevant mixing parameters and the final four-particle mixing matrix (67). From the mixing matrix, the mixing of gluonic contents with the $\eta'$ meson is higher than that of the $\eta$ meson. As shown by Table V and Fig. 3, this make the TFFs $H \to \eta^{(\prime)}(q^2)$ be more sensitive to $H \to \eta(q^2)$, where $H$ represents $B^+$, $D^+$, and $D_s^+$, respectively.

Using the newly derived mixing parameters, we have recalculated the TFFs of $B^+/D^+/D_s^+ \to \eta^{(\prime)}$ by using the QCD LCSRs, where the NLO QCD corrections and the contributions of twist-3 and twist-4 LCDAs have also been included. It has been found that the gluonic contributions are small but sizable to the $B^+/D^+/D_s^+ \to \eta^{(\prime)}$ TFFs. More explicitly, it has shown that their magnitudes are less than 10% for the $B^+/D^+/D_s^+ \to \eta$ TFFs, and less than 20% for the $B^+/D^+/D_s^+ \to \eta'$ TFFs. After extrapolation of those TFFs to whole physical region, our decay widths and decay branching fractions of $B^+/D^+/D_s^+ \to \eta^{(\prime)}\ell^+\nu_\ell$ are consistent with the previous LCSR predictions and experimental results within $1\sigma$. Using the BFTSR estimation of $f_q$ and $f_s$, our determined pseudoscalar glueball mass is $1.509^{+0.133}_{-0.130}$ GeV, which is smaller than that of $X(2370)$. Because the glueball mass increases with the increment of $f_q$ and the decrement of $f_s$, if taking a smaller $f_s$ outside its determined range, we can obtain a larger glueball mass closer to that of $X(2370)$. This may also be due to the fact that only quark-antiquark states were considered in the analysis, while more complex underlying components such as tetraquarks were not included. Some studies have indicated that the $\eta$ state around 1.5 GeV may be dynamically generated [14], and other studies have found that a weak signal of $\eta$ state around 1.5 GeV can be observed when tetraquarks are taken into account [18]. Thus, further studies are still needed to clarify this issue.

## ACKNOWLEDGMENTS

This work was supported in part by the Chongqing Graduate Research and Innovation Foundation under Grants No. CYB23011, No. CYB240057, and No. ydstd1912, and by the Natural Science Foundation of China under Grants No. 12175025, No. 12347101, and No. 12575080.

## DATA AVAILABILITY

The data that support the findings of this article are not publicly available. The data are available from the authors upon reasonable request.

## APPENDIX: THE ELEMENTS OF MASS MATRIX

The mass matrix elements derived from Eq. (16) and related to the mixing angles have the following forms. For $\eta-\eta'-\eta_c$ mixing, the expressions of the mass matrix elements aforementioned are

$$M_{qsc}^{11} = m_\eta^2(c\theta c\theta_i - s\theta s\theta_i c\theta_c)^2 + m_{\eta'}^2(c\theta s\theta_i c\theta_c + s\theta c\theta_i)^2 + m_{\eta_c}^2(s\theta_i)^2(s\theta_c)^2, \tag{A1}$$

$$M_{qsc}^{12} = m_\eta^2((s\theta)^2(c\theta_c)^2 s\theta_i c\theta_i - s\theta c\theta c\theta_c c(2\theta_i) - (c\theta)^2 s\theta_i c\theta_i) + m_{\eta'}^2((c\theta)^2(c\theta_c)^2 s\theta_i c\theta_i + s\theta c\theta c\theta_c c(2\theta_i) - (s\theta)^2 s\theta_i c\theta_i) + m_{\eta_c}^2(s\theta_c)^2 s\theta_i c\theta_i, \tag{A2}$$

$$M_{qsc}^{13} = m_\eta^2 s\theta s\theta_c(s\theta c\theta_c s\theta_i - c\theta c\theta_i) + m_{\eta'}^2 c\theta s\theta_c(c\theta c\theta_c s\theta_i + s\theta c\theta_i) - m_{\eta_c}^2 s\theta_c c\theta_c s\theta_i, \tag{A3}$$

$$M_{qsc}^{21} = M_{qsc}^{12}, \tag{A4}$$

$$M_{qsc}^{22} = m_\eta^2(s\theta c\theta_i c\theta_c + c\theta s\theta_i)^2 + m_{\eta'}^2(c\theta c\theta_i c\theta_c - s\theta s\theta_i)^2 + m_{\eta_c}^2(c\theta_i)^2(s\theta_c)^2, \tag{A5}$$

$$M_{qsc}^{23} = m_\eta^2 s\theta s\theta_c(s\theta c\theta_c c\theta_i + c\theta s\theta_i) + m_{\eta'}^2 c\theta s\theta_c(c\theta c\theta_c c\theta_i - s\theta s\theta_i) - m_{\eta_c}^2 s\theta_c c\theta_c c\theta_i, \tag{A6}$$

$$M_{qsc}^{31} = M_{qsc}^{13}, \tag{A7}$$

$$M_{qsc}^{32} = M_{qsc}^{23}, \tag{A8}$$

$$M_{qsg}^{33} = m_\eta^2(s\theta)^2(s\theta_c)^2 + m_{\eta'}^2(c\theta)^2(s\theta_c)^2 + m_{\eta_c}^2(c\theta_c)^2. \tag{A9}$$

For $\eta-\eta'-G$ mixing, there are equations as follows

$$M_{qsg}^{11} = m_\eta^2(c\theta c\theta_i - s\theta s\theta_i c\phi_G)^2 + m_{\eta'}^2(c\theta s\theta_i c\phi_G + s\theta c\theta_i)^2 + m_G^2(s\theta_i)^2(s\phi_G)^2, \tag{A10}$$

$$M_{qsg}^{12} = -m_\eta^2(s\theta c\theta_i c\phi_G + c\theta s\theta_i)(c\theta c\theta_i - s\theta s\theta_i c\phi_G) + m_{\eta'}^2(c\theta s\theta_i c\phi_G + s\theta c\theta_i)(c\theta c\theta_i c\phi_G - s\theta s\theta_i) + m_G^2 s\theta_i c\theta_i(s\phi_G)^2, \tag{A11}$$

$$M_{qsg}^{21} = M_{qsg}^{12}, \tag{A12}$$

$$M_{qsg}^{22} = m_\eta^2(s\theta c\theta_i c\phi_G + c\theta s\theta_i)^2 + m_{\eta'}^2(c\theta c\theta_i c\phi_G - s\theta s\theta_i)^2 + m_G^2(c\theta_i)^2(s\phi_G)^2, \tag{A13}$$

$$M_{qsg}^{31} = m_\eta^2 s\theta s\phi_G(s\theta s\theta_i c\phi_G - c\theta c\theta_i) + m_{\eta'}^2 c\theta s\phi_G(c\theta s\theta_i c\phi_G + s\theta c\theta_i) - m_G^2 s\theta_i s\phi_G c\phi_G, \tag{A14}$$

$$M_{qsg}^{32} = m_\eta^2 s\theta s\phi_G(s\theta c\theta_i c\phi_G + c\theta s\theta_i) + m_{\eta'}^2 c\theta s\phi_G(c\theta c\theta_i c\phi_G - s\theta s\theta_i) - m_G^2 c\theta_i s\phi_G c\phi_G. \tag{A15}$$

For $\eta - \eta' - G - \eta_c$ mixing, the mass matrix elements related to the mixing angles are

$$M^{11}_{qsgc} = m^2_\eta (c\theta c\theta_i - c\phi_c c\phi_g s\theta s\theta_i)^2 + m^2_{\eta'}(c\theta_i s\theta + c\theta c\phi_c c\phi_g s\theta_i)^2 + m^2_{\eta_c}(s\theta_i)^2(s\phi_c)^2 + m^2_G (c\phi_c)^2 (s\theta_i)^2 (s\phi_g)^2, \tag{A16a}$$

$$\begin{aligned} M^{12}_{qsgc} = {}& m^2_\eta(-c\theta_i c\phi_c c\phi_g s\theta - c\theta s\theta_i)(c\theta c\theta_i - c\phi_c c\phi_g s\theta s\theta_i) + m^2_{\eta'}(c\theta_i s\theta + c\theta c\phi_c c\phi_g s\theta_i)(c\theta c\theta_i c\phi_c c\phi_g - s\theta s\theta_i) \\ & + m^2_{\eta_c} s\theta_i c\theta_i (s\phi_c)^2 + m^2_G c\theta_i (c\phi_c)^2 s\theta_i (s\phi_g)^2, \end{aligned} \tag{A16b}$$

$$\begin{aligned} M^{13}_{qsgc} = {}& -m^2_\eta s\theta(c\theta c\theta_i - c\phi_c c\phi_g s\theta s\theta_i)(c\theta_g c\phi_g s\phi_c + s\theta_g s\phi_g) + m^2_{\eta'} c\theta(c\theta_i s\theta + c\theta c\phi_c c\phi_g s\theta_i)(c\theta_g c\phi_g s\phi_c + s\theta_g s\phi_g) \\ & - m^2_{\eta_c} c\theta_g c\phi_c s\theta_i s\phi_c - m^2_G c\phi_c s\theta_i s\phi_g(c\phi_g s\theta_g - c\theta_g s\phi_c s\phi_g), \end{aligned} \tag{A16c}$$

$$M^{21}_{qsgc} = M^{12}_{qsgc}, \tag{A16d}$$

$$M^{22}_{qsgc} = m^2_\eta(c\theta_i c\phi_c c\phi_g s\theta + c\theta s\theta_i)^2 + m^2_{\eta'}(c\theta c\theta_i c\phi_c c\phi_g - s\theta s\theta_i)^2 + m^2_{\eta_c}(c\theta_i)^2(s\phi_c)^2 + m^2_G (c\theta_i)^2 (c\phi_c)^2 (s\phi_g)^2, \tag{A16e}$$

$$\begin{aligned} M^{23}_{qsgc} = {}& m^2_\eta s\theta(c\theta_i c\phi_c c\phi_g s\theta + c\theta s\theta_i)(c\theta_g c\phi_g s\phi_c + s\theta_g s\phi_g) + m^2_{\eta'} c\theta(c\theta c\theta_i c\phi_c c\phi_g - s\theta s\theta_i)(c\theta_g c\phi_g s\phi_c + s\theta_g s\phi_g) \\ & - m^2_{\eta_c} c\theta_g c\theta_i c\phi_c s\phi_c - m^2_G c\theta_i c\phi_c s\phi_g(c\phi_g s\theta_g - c\theta_g s\phi_c s\phi_g), \end{aligned} \tag{A16f}$$

$$\begin{aligned} M^{31}_{qsgc} = {}& m^2_\eta s\theta(c\theta c\theta_i - c\phi_c c\phi_g s\theta s\theta_i)(c\phi_g s\theta_g s\phi_c - c\theta_g s\phi_g) + m^2_{\eta'} c\theta(c\theta_i s\theta + c\theta c\phi_c c\phi_g s\theta_i)(c\theta_g s\phi_g - c\phi_g s\theta_g s\phi_c) \\ & + m^2_{\eta_c} c\phi_c s\theta_g s\theta_i s\phi_c - m^2_G c\phi_c s\theta_i s\phi_g(c\theta_g c\phi_g + s\theta_g s\phi_c s\phi_g), \end{aligned} \tag{A16g}$$

$$\begin{aligned} M^{32}_{qsgc} = {}& m^2_\eta s\theta(c\theta_i c\phi_c c\phi_g s\theta + c\theta s\theta_i)(c\theta_g s\phi_g - c\phi_g s\theta_g s\phi_c) + m^2_{\eta'} c\theta(c\theta c\theta_i c\phi_c c\phi_g - s\theta s\theta_i)(c\theta_g s\phi_g - c\phi_g s\theta_g s\phi_c) \\ & + m^2_{\eta_c} c\theta_i c\phi_c s\theta_g s\phi_c - m^2_G c\theta_i c\phi_c s\phi_g(c\theta_g c\phi_g + s\theta_g s\phi_c s\phi_g), \end{aligned} \tag{A16h}$$

$$\begin{aligned} M^{33}_{qsgc} = {}& m^2_\eta (s\theta)^2(c\theta_g s\phi_g - c\phi_g s\theta_g s\phi_c)(c\theta_g c\phi_g s\phi_c + s\theta_g s\phi_g) + m^2_{\eta'}(c\theta)^2(c\theta_g s\phi_g - c\phi_g s\theta_g s\phi_c)(c\theta_g c\phi_g s\phi_c + s\theta_g s\phi_g) \\ & - m^2_{\eta_c} s\theta_g c\theta_g (c\phi_c)^2 + m^2_G (c\phi_g s\theta_g - c\theta_g s\phi_c s\phi_g)(c\theta_g c\phi_g + s\theta_g s\phi_c s\phi_g), \end{aligned} \tag{A16i}$$

$$M^{41}_{qsgc} = M^{13}_{qsgc}, \tag{A16j}$$

$$M^{42}_{qsgc} = M^{23}_{qsgc}, \tag{A16k}$$

$$\begin{aligned} M^{43}_{qsgc} = {}& m^2_\eta (s\theta)^2(c\theta_g c\phi_g s\phi_c + s\theta_g s\phi_g)^2 + m^2_{\eta'}(c\theta)^2(c\theta_g c\phi_g s\phi_c + s\theta_g s\phi_g)^2 + m^2_{\eta_c}(c\theta_g)^2(c\phi_c)^2 \\ & + m^2_G (c\phi_g s\theta_g - c\theta_g s\phi_c s\phi_g)^2. \end{aligned} \tag{A16l}$$

---